\documentclass[english,aps,prl,twocolumn,superscriptaddress]{revtex4}
\usepackage{amsmath}
\usepackage{amssymb}
\usepackage[dvipdfmx]{graphicx}
\usepackage{bm}
\usepackage{color}
\usepackage{comment}
\makeatletter
\usepackage[colorlinks=true,urlcolor=blue,citecolor=blue,linkcolor=blue,breaklinks=true,dvipdfmx]{hyperref}
\makeatother
\begin{document}

\title{Why the critical temperature of high-$T_{\rm c}$ cuprate superconductors is so low:\\
       The importance of the dynamical vertex structure }

\author{Motoharu Kitatani}
\affiliation{Institute of Solid State Physics, Vienna University of Technology, A-1040 Vienna, Austria}
\affiliation{Department of Physics, University of Tokyo, Hongo, Tokyo 113-0033, Japan}

\author{Thomas Sch\"{a}fer}
\affiliation{Institute of Solid State Physics, Vienna University of Technology, A-1040 Vienna, Austria}
\affiliation{Coll\`{e}ge de France, 11 place Marcelin Berthelot, 75005 Paris, France}
\affiliation{Centre de Physique Th\'{e}orique, \'{E}cole Polytechnique, CNRS, route de Saclay, 91128 Palaiseau, France}

\author{Hideo Aoki}
\affiliation{Department of Physics, University of Tokyo, Hongo, Tokyo 113-0033, Japan}
\affiliation{Electronics and Photonics Research Institute,
Advanced Industrial Science and Technology (AIST), Tsukuba, Ibaraki 305-8568, Japan}

\author{Karsten Held}
\affiliation{Institute of Solid State Physics, Vienna University of Technology, A-1040 Vienna, Austria}

\date{\today}
\begin{abstract}
To fathom the mechanism of high-temperature ($T_{\rm c}$) superconductivity,
the dynamical vertex approximation (D$\Gamma$A) is evoked for 
the two-dimensional repulsive Hubbard model.  
After showing that our results well reproduce the cuprate phase diagram with a
reasonable $T_{\rm c}$ and dome structure, 
we keep track of the scattering processes that 
primarily affect $T_{\rm c}$.  We 
find that local {\it particle-particle} diagrams significantly 
screen the bare interaction at low frequencies, 
which in turn suppresses antiferromagnetic spin fluctuations and hence the pairing interaction. 
Thus we identify dynamical vertex corrections as one of the main 
oppressors of $T_{\rm c}$, which may provide a hint toward higher $T_{\rm c}$'s.
\end{abstract}

\maketitle

{\em Introduction.}
More than three decades after the discovery of the 
high-$T_{\rm c}$ cuprate superconductors \cite{Bednorz1986}, the quest
for higher (or even room-temperature) $T_{\rm c}$ superconductors remains one of the biggest challenges in solid-state physics. 
Despite intensive efforts, 
we are still stuck with  $T_{\rm c} \lesssim 130 \,$K \cite{Schilling1993}. 
Nonetheless the cuprates do remain the arguably most promising 
material class, at least at ambient pressure \cite{Drozdov2015}. 

In this arena, theoretical estimations of $T_{\rm c}$, 
specifically identifying the reason why it is so low 
(as compared with the starting electronic energy scales 
of $\sim$ eV), should be imperative if one wants to 
possibly enhance $T_{\rm c}$.  Through many theories 
proposed and intensively debated, 
it has become clear that superconductivity in the cuprates is interlinked with electronic correlations, 
which are considered to mediate the pairing through spin fluctuations \cite{Scalapino12}. 
The simplest and most widely used model for cuprates is the repulsive Hubbard model on a square lattice, where a formidable problem is that the scale of 
$T_{\rm c}$ is orders of magnitude smaller than the 
Hubbard interaction $U$ and the hopping amplitude $t$, 
which has been a key question from the early stage of high-$T_{\rm c}$ studies \cite{Lee}.
Various approaches have been employed to attack the problem, see e.g. Refs. \cite{Kyung2003,Maier2005,PhysRevLett.116.057003,Lee2006,Hafermann2009a,Metzner2012,Otsuki2014,Kitatani2015,Vucicevic2017}.
Thus, while the conventional phonon-mediated superconductors can now be accurately captured by
density functional theory for superconductors (SCDFT) \cite{SCDFT1,SCDFT2}, a full understanding of 
$T_{\rm c}$ in the Hubbard model has yet to come.  
One inherent reason for the low $T_{\rm c}$ is the  $d$-wave symmetry of the gap function arising from the local repulsion. 
%The gap function has nodes on the Fermi surface which intervene the energy gains as 
%compared with the simple $s$-wave pairing.

A possibly essential mechanism that reduces $T_{\rm c}$ comes from vertex corrections.  
Migdal's theorem \cite{Migdal1958}, which works so 
nicely for phonon-mediated pairing, is no longer 
applicable to unconventional superconductivity 
due to the electron correlation. For strongly correlated systems,
we should in fact 
expect vertex corrections to be a major player, affecting $T_{\rm c}$ and changing it with respect to simpler (e.g., mean-field-like) treatments \cite{PhysRevLett.112.187001,PhysRevB.96.235110}.

Thanks to recent extensions of the dynamical mean-field theory (DMFT) \cite{Metzner1989,Georges1992a,Georges1996}, 
specifically the dynamical vertex approximation (D$\Gamma$A) \cite{Toschi2007,Kusunose2006,Slezak2009,Katanin2009}, 
the dual-fermion \cite{Rubtsov2008} 
and other related approaches \cite{Rohringer2013,Taranto2014,Ayral2015,Li2015,Ayral2016a,Ayral2016,Vucicevic2017}, such  
vertex corrections can now be studied for strong correlations; see \cite{RMPVertex} for a review.
Owing to this development we now understand the (local) 
vertex structures much better \cite{RMPVertex,Rohringer2012,Rohringer2013a,Wentzell2016}, e.g., 
how they affect the spectral function and lead to pseudogaps in the normal phase 
\cite{Katanin2009,Rubtsov2009,Schaefer2015-2,Schaefer2015-3,Taranto2014,Ayral2015,Gunnarsson2015,Pudleiner2016,gunnarsson2017complementary}. 
This now puts us in a position to shed light on the impact of dynamical vertex corrections
on superconductivity. 

\begin{figure}[t]
\begin{centering}
\includegraphics[width=1.0\columnwidth]{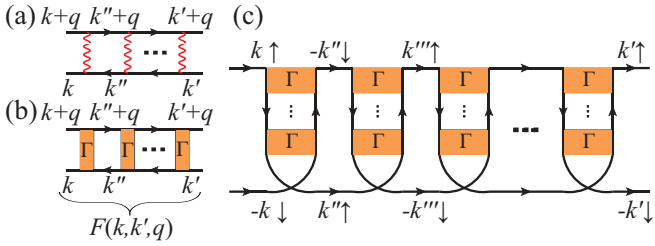}
\par\end{centering}
\caption{(Color online)
(a) Antiferromagnetic spin fluctuations captured for weak interaction $U$ (wiggled line) in terms of particle-hole ladder diagrams (solid line: Green's function). 
(b) D$\Gamma$A diagrams
describe similar spin fluctuations but now for strong correlation, with ladders in terms  of the local $\Gamma$ which is 
non-perturbative and frequency-dependent \cite{RMPVertex} instead of $U$.
(c) The spin fluctuations can act, in turn, as a pairing glue for superconductivity in the particle-particle channel (an exemplary diagram is shown).
}
\label{fig:intro}
\end{figure}

In this paper, we analyze how vertex corrections affect $T_{\rm c}$.
%To this end, we first extend the D${\rm \Gamma}$A 
%formalism to superconductivity.  
We find that the dynamical structure (frequency dependence) 
of the vertex, $\Gamma(\nu,\nu^{\prime},\omega)$,  
is actually essential for estimating $T_{\rm c}$. 
Note that  $\Gamma$ is non-perturbative; 
it sums up the local contribution of all Feynman diagrams (to all orders in the interaction) connecting two incoming and two outgoing particles. Physics of strongly correlated electrons such as the quasiparticle renormalization and the formation of Hubbard bands are hence encoded in  $\Gamma$. On top of such correlations already included in DMFT, the D$\Gamma$A  further incorporates non-local correlations, in particular spin and superconducting fluctuations, see Fig.~\ref{fig:intro} and \cite{RMPVertex}. 
The present results show that the dynamics of $\Gamma$, which turns out to reduce 
the pairing interaction in a low-frequency regime, suppresses $T_{\rm c}$ by one order of magnitude. 
We unravel the physical origin in the relevant dynamical vertex structure, as it is passed 
from the local vertex to the magnetic vertex describing antiferromagnetic spin fluctuations and, eventually, to the pairing interaction. 

{\em Model and methods.}
We consider the two-dimensional single-orbital Hubbard model,
\begin{equation}
{\cal H} = \sum_{\bm{k},\sigma} \epsilon(\bm{k})c^{\dag}_{\bm{k},\sigma}c_{\bm{k},\sigma}+U\sum_{i} {\hat n}_{i \uparrow}{\hat n}_{i \downarrow} \; ,
\end{equation}
where $c^{\dag}_{\bm{k},\sigma}$ ($c^{\phantom{\dag}}_{\bm{k},\sigma}$) creates (annihilates) an electron 
with spin $\sigma=\uparrow, \downarrow$ and wave-vector $\bm{k}$, $U$ is the on-site Coulomb repulsion, and 
${\hat n}_{i \sigma} \equiv c^{\dag}_{i \sigma}c_{i \sigma}$. The two-dimensional band dispersion is given by
$\epsilon(\bm{k}) = -2t( {\rm cos}k_x+{\rm cos}k_y ) -4t^{\prime}{\rm cos}k_x{\rm cos}k_y	-2t^{\prime \prime}( {\rm cos}2k_x+{\rm cos}2k_y )$
with  $t$, $t^{\prime}$, and $t^{\prime \prime}$ being the nearest, second, and third neighbor hoppings, respectively. 
We consider two sets of hopping parameters:
(a) $t^{\prime}=t^{\prime \prime}=0$, and
(b) $t^{\prime}/t=-0.20, t^{\prime \prime}/t=0.16$ which corresponds 
to the band-structure of HgBa$_2$CuO$_{4+\delta}$ \cite{nishiguchi2013,nishiguchiphd}.

We adopt the D${\rm \Gamma}$A 
as a method that incorporates non-local correlations beyond the local correlations treated in DMFT. 
%Here we extend it for describing superconductivity.
In the D${\rm \Gamma}$A \cite{Toschi2007,Katanin2009,RMPVertex}, 
the local two-particle vertex $\Gamma$ that is irreducible in the particle-hole channel
is calculated from a DMFT impurity problem. 
We employ the exact diagonalization  as an impurity solver to this end, 
but also checked against  quantum Monte Carlo simulations \cite{CTHYB1,CTHYB2,Parragh2012,W2DYN}, 
%for some parameters
see Supplemental Material \cite{suppl}. 

From $\Gamma_{\sigma \sigma^{\prime}}(\nu,\nu^{\prime},\omega)$, 
the non-local vertex $F_{\sigma\sigma^{\prime}}(k,k^{\prime},q)$, 
which describes, among others, longitudinal and transversal spin-fluctuations, 
is obtained via the Bethe-Salpeter equation 
in the vertical particle-hole channel [as visualized in Fig.~\ref{fig:intro}~(b)] 
and transversal particle-hole channel [not shown]. 
$F_{\sigma\sigma'}(k,k^{\prime},q)$ depends on the spin $(\sigma,\sigma^{\prime})$, 
two fermionic $(k,k^{\prime})$ and one bosonic $(q)$ 
four-vectors consisting
of momentum and Matsubara frequency, i.e. $k=(\bm{k},\nu)$. 
From $F$, the  D${\rm \Gamma}$A self-energy $\Sigma(k)$ is in turn computed via the Schwinger-Dyson equation \cite{RMPVertex};
 spin fluctuations included in $\Sigma(k)$ give rise to  a pseudogap in the  non-local Green's function $G(k)$  \cite{Katanin2009,Schaefer2015-3,suppl}.  

For studying superconductivity, we extend here the existing D${\rm \Gamma}$A treatment. 
That is, we extract, from $F$, the particle-particle irreducible vertex $\Gamma_{\rm pp}(k,k^{\prime},q=0) \equiv F(k^{\prime},-k,k-k^{\prime}) - \Phi_{\rm pp}(\nu,\nu^{\prime},\omega=0)$ (with four-vector 
in particle-particle convention).  
Here $\Phi_{\rm pp}$ is defined as the {\em local} reducible vertex  diagrams in the particle-particle channel, 
which are included in $F$ but need to be subtracted to obtain the 
%${\rm pp}$-irreducible vertex  
$\Gamma_{{\rm pp},\bm{Q}=\bm{k}-\bm{k}^{\prime}}(\nu,\nu^{\prime})$,
%$(k,k^{\prime},q=0)$ 
see \cite{suppl} for details. 
The vertex $\Gamma_{\rm pp}$ contains spin fluctuations as a pairing glue, and we can now 
insert it into the  particle-particle ladder  [as illustrated in Fig.~\ref{fig:intro}~(c) for selected diagrams]. 
For evaluating this ladder, we use the linearized gap (Eliashberg) equation \cite{eliashberg}:
\begin{equation}
\lambda \Delta(k) = -\frac{1}{\beta N_{\bm{k}}} \sum_{k^{\prime}}
\Gamma_{\rm pp}(k,k^{\prime},q=0)G(k^{\prime})G(-k^{\prime})\Delta(k^{\prime}).
\label{Eq:Eliashberg}
\end{equation}
Here, $\Delta(k)$ is the anomalous self-energy, $\lambda$ the superconducting eigenvalue with $\lambda\rightarrow 1$ 
signaling an instability toward superconductivity \cite{Footnote1}, $\beta=1/T$ the inverse temperature, and $N_{\bm{k}}$ the number of $\bm{k}$ points.

\begin{figure}[t]
\begin{centering}
\includegraphics[width=1\columnwidth]{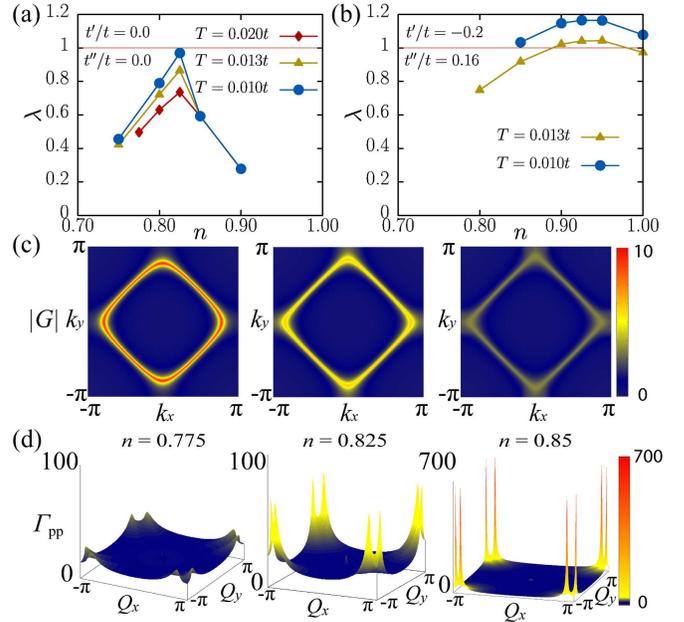}
\par\end{centering}
\caption{(Color online) $d$-wave eigenvalue $\lambda$ against the band filling $n$ for  $U=6t, T/t = 0.010, 0.013, 0.020$ with (a) $t^{\prime}=t^{\prime\prime}=0$ and (b) $t^{\prime}/t=-0.20, t^{\prime \prime}/t=0.16$.
(c,d) Momentum dependence of the Green's function 
$|G(\pi/\beta,{\bm{k}})|$ (c) and   
the pairing interaction vertex $\Gamma_{{\rm pp},{\bm Q}}$ (d) 
for $n=0.775$ (overdoped), $0.825$ (optimally doped), and $0.85$ (underdoped), at $T/t=0.02$ with other parameters as in (a). 
In (d) we specifically 
display the ${\bm Q}$ dependence of the pairing vertex 
$[\Gamma_{{\rm pp},{\bm Q}}(\pi/\beta,\pi/\beta)+\Gamma_{{\rm pp},{\bm Q}}(-\pi/\beta,\pi/\beta)]/2$,
which is symmetrized for d-wave (singlet, even-frequency) pairing.}
\label{fig:phase}
\end{figure}

{\em Size and dome shape of $T_{\rm c}$.}
We first show the superconducting
eigenvalue $\lambda$  for the two sets of hopping parameters  in Fig.~\ref{fig:phase}~(a) and (b), respectively. 
In the doping region in Fig.~\ref{fig:phase}, the $d$-wave 
has the largest $\lambda$, 
while antiferromagnetic fluctuations become dominant
close to half-filling, c.f. Refs.~\cite{Katanin2009,Schaefer2016}. % for Fig.~\ref{fig:phase}~(a).
A superconducting instability  ($\lambda\rightarrow 1$) is found for $T_{\rm c}\lesssim 0.01t$ in  Fig.~\ref{fig:phase}~(a) and for $T_{\rm c} \approx 0.015t$ in Fig.~\ref{fig:phase}~(b).

The results well reproduce the phase diagram of the cuprates with
a  dome structure and peaks that amount to 
$T_{\rm c} \approx 50-80$ K around $n = 0.80 - 0.95$ 
if we take a typical $t\approx 0.45$ eV \cite{sakakibara2010}.
We can explain the physical origin of the $T_c$ dome 
as follows: antiferromagnetic spin fluctuations that mediate the pairing 
become stronger [i.e., $\Gamma_{\rm pp}$ in Fig.~\ref{fig:phase}~(d) 
increases] toward half-filling, while close to half-filling 
the self-energy blows up and damps $|G(k)|$ in Fig.~\ref{fig:phase}~(c). 
The latter eventually leads  to a pseudogap at smaller dopings within the central peak in a three-peak spectrum, see Supplemental Material \cite{suppl}.
Thus the dome appears as a consequence of two opposing factors: $\Gamma_{\rm pp}$ and $G(k)$
 in the gap equation~(\ref{Eq:Eliashberg}). 
We can see in  Fig.~\ref{fig:phase}~(d) that $\Gamma_{\rm pp}$ is sharply peaked at around ${\bm Q}=(\pm\pi,\pm\pi)$ (with some offset and splitting because of incommensurability), leading to a $d$-wave $\Delta(k)$ in  Eq.~(\ref{Eq:Eliashberg}).
Let us note that a superconducting dome has also been reported in e.g.\ \cite{Kyung2003,Maier2005,Kitatani2015,Vucicevic2017}, 
but not in the dual-fermion approach \cite{Hafermann2009a,Otsuki2014}.

\begin{figure}[t]
\begin{centering}
\includegraphics[width=1\columnwidth]{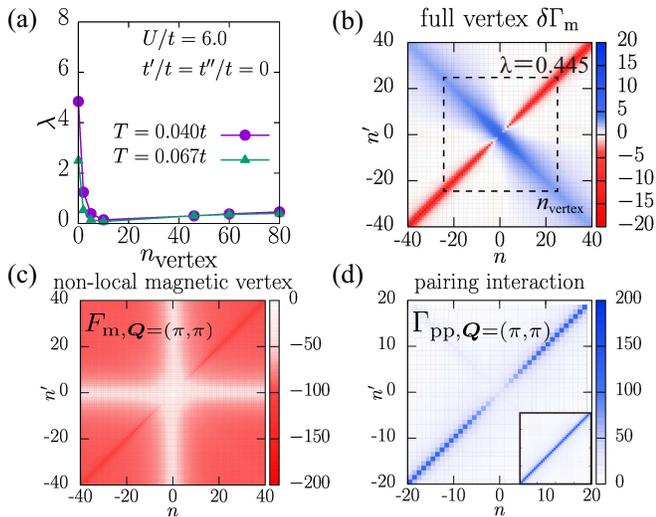}
\par\end{centering}
\caption{(Color online)
(a) Eigenvalue $\lambda$ against the frequency range $n_{\rm vertex}$ [exemplified in (b) as a dashed line], over which the local vertex correction $\delta\Gamma_{\rm m}$  is considered, 
for $U=6t, t^{\prime}=t^{\prime \prime}=0, n=0.825$ and $T/t=0.040, 0.067$. 
(b,c,d) Dynamical vertex structure of (b) the local vertex correction 
$\delta \Gamma_{\rm m}(\nu_n,\nu_{n^{\prime}},\omega=0)$ in the magnetic channel  relative to $-U$, 
(c) the non-local vertex in the magnetic channel, $F_{{\rm m},{\bm Q}=(\pi,\pi)}(\nu_n,\nu_{n^{\prime}},\omega=0)$, 
and (d) the pairing interaction $\Gamma_{{\rm pp},{\bm Q}=(\pi,\pi)}(\nu_n,\nu_{n^{\prime}},\omega=0)$, 
for the same $U, t^{\prime},t^{\prime \prime}, n$ with $T/t= 0.067$ here.
The inset in (d) shows a typical structure of $\Gamma_{\rm pp}$ in  mean-field-like approaches.
}
\label{fig:vertex-structure}
\end{figure}

\begin{figure}[t]
\begin{centering}
\includegraphics[width=1\columnwidth]{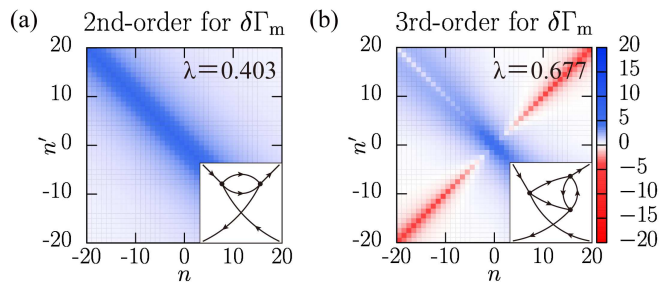}
\par\end{centering}
\caption{(Color online)
Frequency structure of the local vertex correction, $\delta\Gamma_{\rm m}(\nu_n,\nu_{n^{\prime}},\omega=0)\equiv\Gamma_{\rm m}(\nu_n,\nu_{n^{\prime}},\omega=0)-(-U)$, for 
(a) second-order, or (b) third-order perturbation theory for
$U=6t, t^{\prime}=t^{\prime \prime}=0, n=0.825$, and $T/t=0.067$. In 
each panel, the inset 
shows a typical diagram taken into account, and the 
corresponding eigenvalue $\lambda$ is indicated. 
}
\label{fig:pert}
\end{figure}

{\em Importance of the dynamical vertex structure.}
Let us now look into the structure of the vertex 
$\Gamma(\nu,\nu^{\prime},\omega=0)$ against frequencies $\nu,\nu^{\prime}$. 
%Indeed, this turns out to be a key for obtaining the correct $T_{\rm c}$.
As we shall see below,  if we start from a 
mean-field or random phase approximation (RPA)-like treatment, where 
$\Gamma_{\rm m}(\nu,\nu^{\prime},\omega)=\! \Gamma_{\uparrow \uparrow} \!-\! \Gamma_{\uparrow \downarrow}$, is replaced with the bare $-U$ in the Bethe-Salpeter ladder \cite{FLEX+DMFT}, this 
would yield stronger spin fluctuations and overestimate  $T_{\rm c}$ by 
an order of magnitude. 

We can elucidate this point in an energy-resolved fashion by taking 
the local irreducible vertex in the magnetic channel $\Gamma_{\rm m}(\nu,\nu^{\prime},\omega)$
only up to a frequency  $n_{\rm vertex}$, with the bare $(-U)$ 
adopted outside this range \cite{freqrange} (pictorially this means taking Fig.~\ref{fig:intro}~(a) instead of  Fig.~\ref{fig:intro}~(b) for large frequencies). 
In Fig.~\ref{fig:vertex-structure}~(a), 
we plot the eigenvalue $\lambda$ against $n_{\rm vertex}$ \cite{note-nvertex}.
As the region $n_{\rm vertex}$ in which we take the dynamical vertex is widened, 
$\lambda$ is seen to dramatically decrease, 
already when a few  frequencies are taken into account.
%If we plot in Fig.~\ref{fig:vertex-structure}~(a) 
%the eigenvalue $\lambda$ against $n_{\rm vertex}$ \cite{note-nvertex}, 
%$\lambda$ is seen to dramatically decrease, 
%already when a few  frequencies are taken into account, 
%as $n_{\rm vertex}$ is widened.  
This signifies that the {\it low-frequency part} 
of the dynamical $\Gamma_{\rm m}$  is quite important.

Figure~\ref{fig:vertex-structure}~(b) displays the deviation, 
$\delta \Gamma_{\rm m}
(\nu_n,\nu_{n^{\prime}},\omega=0)
\equiv \Gamma_{\rm m}
(\nu_n,\nu_{n^{\prime}},\omega=0)
-(-U)$, of the local $\Gamma_m$ from $-U$.
$\Gamma_{\rm m}$ is indeed prominently reduced 
(dark blue $\delta \Gamma_{\rm m}$) 
at small frequencies $\nu_n,\nu_{n^{\prime}}\sim 0$, as well as along the diagonal $\nu_n=-\nu_{n^{\prime}}$.
%This suppressed vertex  $\Gamma_{\rm m}$ 
% in turn is at the origin of the reduced eigenvalues in Fig.~\ref{fig:vertex-structure}~(a) \cite{note-nvertex}.

The non-local magnetic vertex 
$F_{\rm m}$ [Fig.~\ref{fig:vertex-structure}~(c)] and the pairing interaction 
$\Gamma_{\rm pp}$ [Fig.~\ref{fig:vertex-structure}~(d)] 
inherit similar dynamical structures from  the local  $\Gamma_{\rm m}$.
%:
%i.e., reductions at small frequencies near the Fermi level. 
This comes from the Bethe-Salpeter equation
$F_{\rm m} = \Gamma_{\rm m} - \Gamma_{\rm m}\chi_0 F_{\rm m}$ with $\chi_0$ being the bare bubble susceptibility, and from $\Gamma_{\rm pp}=F-\Phi_{\rm pp}$, respectively; 
for a more extensive discussion, see Supplemental Material \cite{suppl}. 
Without the vertex $\delta \Gamma_{\rm m}$,
$F_{\rm m}$ depends on the spin susceptibility through $F_{\rm m}= -U - U^2 \chi_{\rm m}(\omega)$ \cite{suppl} and hence only on $\omega$, which corresponds to the red background in   Fig.~\ref{fig:vertex-structure}~(c). As a consequence, the pairing vertex $\Gamma_{\rm pp}$ depends only on the difference of two frequencies $\nu_n,\nu_{n^{\prime}}$ [inset of Fig.~\ref{fig:vertex-structure}~(d)].
%With the vertex correction $\delta \Gamma_{\rm m}$ in Fig.~\ref{fig:vertex-structure}~(c), 
%as well as the red background reflecting the susceptibility $\chi_{\rm m}(\omega)$ 
%(which depends only on $\omega$),
%there is a structure depending on ($\nu_n$,$\nu_{n^{\prime}}$), 
%which also leads to the reduction of the pairing interaction $\Gamma_{\rm pp}$ at low frequencies in Fig.~\ref{fig:vertex-structure}~(d).
%With the vertex correction $\delta \Gamma_{\rm m}$ in Fig.~\ref{fig:vertex-structure}~(c), 
%the susceptibility $\chi_{\rm m}(\omega)$ is reduced from the 
%background (paler color) for small $\nu_n$,$\nu_{n^{\prime}}$,
%which in turn leads the reduction of the pairing interaction $\Gamma_{\rm pp}$ at low frequencies in Fig.~\ref{fig:vertex-structure}~(d).
With the suppressed  $\Gamma_{\rm pp}$, the  Eliashberg Eq.~(\ref{Eq:Eliashberg}) finally leads to a reduced  $\lambda$ and $T_{\rm c}$.
Thus we have traced that the local vertex corrections are responsible for the reduction of $T_{\rm c}$, where 
an important message is that their {\it dynamical} structures has to be examined. 
Indeed, mean-field-like (e.g., paramagnon-exchange) picture cannot describe the frequency structure in Fig.~\ref{fig:vertex-structure}~(d)
even if we consider vertex correction effects on the susceptibility $\chi_{\rm m}(\omega)$.
%This
%goes {\it beyond} the magnetic susceptibility that is a quantity summed over
%$\nu_n$ and $\nu_{n^{\prime}}$
%, see Supplemental Material \cite{suppl}

{\em Physics behind suppression of $\Gamma_{\rm m}$.}
Having identified the suppression of the local $\Gamma_{\rm m}$ 
as the key ingredient for low $T_{\rm c}$'s, 
we can now pin-point which physical processes are at its origin. 
In Fig.~\ref{fig:pert}  we show the contributions to $\delta \Gamma_{\rm m}$  in (a) second-order and (b) third-order perturbation theory, where we show a typical diagram 
along with the eigenvalue $\lambda$ estimated in 
D$\Gamma$A when $\Gamma_{\rm m}$ is replaced by the displayed 
local vertex \cite{note-nvertex}.  
When the bare value ($-U$; $\delta \Gamma_{\rm m}=0$) is used instead of the full  $\Gamma_{\rm m}$,
 $\lambda$ is enhanced dramatically from the correct value 0.45 to 2.49  for $T/t=0.067$
($T_{\rm c}$ increases correspondingly from 0.01t to 0.13t). We can see that 
most of the dynamical effect is already included in the second-order particle-particle diagram in Fig.~\ref{fig:pert}(a), 
which compensates the bare contribution ($-U$) for $\nu_n\approx -\nu_{n^{\prime}}$, and
strongly reduces $\lambda$ back to 0.40.  
Third-order diagrams 
in  Fig.~\ref{fig:pert}~(b) slightly enhance $\lambda$, 
and already resemble the full vertex qualitatively. 
Thus the second-order particle-particle diagrams in Fig.~\ref{fig:pert}~(a) constitute by far the major process for the suppression of the $\lambda$. 

\begin{figure}[t]
\begin{centering}
\includegraphics[width=1.0\columnwidth]{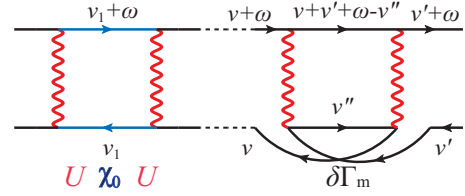}
\par\end{centering}
\caption{(Color online)
Typical diagrams that contribute to the magnetic vertex $F_{\rm m}$ in the Bethe-Salpeter ladder. 
Left: a typical diagram as in RPA with $U$ (red wavy line) as an irreducible building block, 
connected by $\chi_0$ (with two Green's functions having
 fermionic frequencies $\nu_1$ and $\nu_1+\omega$). 
Right: local (second-order) vertex correction $\delta\Gamma_{\rm m}$ 
with a {\em particle-particle} bubble. Such terms are particle-hole irreducible, hence need to be inserted in the Bethe-Salpeter ladder 
for spin fluctuations. 
They lead to the suppression of  $\Gamma_{\rm m}$ in Fig.~\ref{fig:pert}, and 
through the whole ladder 
suppress  $F_{\rm m}$ and $\Gamma_{\rm pp}$ in Fig.~\ref{fig:vertex-structure} (c,d).
}
\label{fig:diags}
\end{figure}

Hence it is worthwhile to look into this second-order contribution in more detail.  The local irreducible vertex $\Gamma_{\rm m}$ is the building block for the
non-local particle-hole ladder that leads to  magnetic fluctuations as visualized in Fig.~\ref{fig:intro}~(b). 
If we take  $\Gamma_{\rm m}(\nu,\nu^{\prime},\omega)=-U$ as in Fig.~\ref{fig:intro}~(a) or the left part of Fig.~\ref{fig:diags}, 
we obtain the standard RPA  with a 
Stoner-enhanced spin susceptibility,
\begin{equation}
\chi = \chi_0/(1-U\chi_0)=\chi_0+\chi_0U\chi_0+\chi_0U\chi_0U\chi_0 \ldots \; .
\label{Eq:RPA}
\end{equation} 
While all the terms enhance the susceptibility in this geometric series in 
$U$, local vertex corrections do need to be included in the particle-hole ladder with $\Gamma_{\rm m}(\nu,\nu^{\prime},\omega)=-U+\delta\Gamma_{\rm m}(\nu,\nu^{\prime},\omega)$ as a building block.
In Fig.~\ref{fig:pert}~(a) we have identified  the second-order particle-particle contribution to $\delta\Gamma_{\rm m}>0$ to be most important for suppressing  antiferromagnetic spin fluctuations, and the inclusion of such a contribution $\delta\Gamma_{\rm m}$ in the  ladder series for $F_m$ is visualized  in the right part of Fig.~\ref{fig:diags}.

The difference from the RPA ladder comprising $-U$ and {\em particle-hole} bubbles (left block in Fig.~\ref{fig:diags})  is that  $\delta\Gamma_{\rm m}>0$ 
has (in the second order) two $-U$'s and a local {\em particle-particle} bubble (right  block in  Fig.~\ref{fig:diags}).  This bubble, being a particle-particle bubble, 
depends on the frequency combination $\nu+\nu^{\prime}+\omega$ 
rather than on $\omega$ alone as in particle-hole bubbles. 
This, first of all, gives the pronounced frequency structure of $\delta\Gamma_{\rm m}$ in Fig.~\ref{fig:pert} (a). 
Since Fig.~\ref{fig:diags} shows typical diagrams that contribute to $F_{\rm m}$, we can also see that, with a 
 $\delta\Gamma_{\rm m}$ located at an end of the ladder, 
$F_{\rm m}$ and $\Gamma_{\rm pp}=F-\Phi_{\rm pp}$ inherits a similar frequency structure as in Figs.~\ref{fig:vertex-structure} (c,d);
see Supplemental Material \cite{suppl} for a general explanation.

Second, at its maximum $(\omega=0, \nu^{\prime}=-\nu)$, the particle-particle bubble 
$\sum_{\nu^{\prime\prime}} G(\nu^{\prime\prime}) G(-\nu^{\prime\prime}) \!=\! \sum_{\nu^{\prime\prime}} G(\nu^{\prime\prime}) G^*(\nu^{\prime\prime})$,
has a sign opposite to the particle-hole bubble  $\sum_{\nu^{\prime\prime}} G(\nu^{\prime\prime}) G(\nu^{\prime\prime})$, 
because the biggest contribution comes from ${\rm Im }\, G(\nu)$.
Hence $\delta\Gamma_{\rm m}$ partially compensates 
the second-order RPA contribution, $U\chi_0U$ in Eq.~(\ref{Eq:RPA}).
This is the reason why $\delta \Gamma_{\rm m}$  reduces the bare $U$,
whereas the RPA ladders Eq.~(\ref{Eq:RPA}) enhances it.

Now we are in a position to finally grasp a physical picture: 
while a repulsive interaction can give rise to 
a spin-fluctuation mediated attraction through the particle-hole channel, 
a local repulsive interaction $U$  always 
leads to a repulsion  between two particles in the particle-particle channel, too.
As we have seen in Fig.~\ref{fig:diags}, this repulsion in the particle-particle channel 
 reduces the antiferromagnetic spin fluctuations, 
albeit only for certain  frequency combinations. With reduced antiferromagnetic spin fluctuations, superconductivity is suppressed.

{\em Conclusion and outlook.}
We have extended the D$\Gamma$A 
formalism for studying superconductivity in the repulsive 
Hubbard model on a square lattice.  
Our results well reproduce the superconducting dome and typical values of $T_{\rm c}\approx 50-80\;$K  for the cuprates. We have pinpointed the  importance of dynamical vertex corrections.
That is, $T_{\rm c}$ would be around room temperature 
if the pairing interaction was built from a ladder with the bare interaction $U$. 
However, local vertex corrections give rise to a pronounced frequency structure 
accompanied with a suppression (screening) of the irreducible magnetic vertex  $\Gamma_{\rm m}$ 
(i.e., the effective  interaction in the magnetic channel). 
This in turn suppresses antiferromagnetic spin fluctuations and the pairing glue  ($\Gamma_{\rm pp}$) for superconductivity in the particle-particle channel. 

Thus local particle-particle fluctuations are at the origin of the 
suppression of $\Gamma_{\rm m}$, so that it is intriguing to ask: can one possibly evade this oppressor of $T_{\rm c}$?  This is not simple:  As 
the leading correction that reduces the  bare interaction $U$ is $\sim U^2$,
the suppression becomes smaller for weaker Coulomb interactions, but so do the antiferromagnetic spin fluctuations. 
Local particle-particle fluctuations can be suppressed by disorder or a magnetic field \cite{Altshuler1985}, 
but this would degrade the non-local particle-particle (superconducting) fluctuations, too.
One way-out might be to 
exploit the characteristic frequency structure of 
$\Gamma_{\rm m}$, possibly in combination with a frequency-dependent (local) interaction, 
which may originate from off-site (extended Hubbard) 
interactions as described in dual-boson \cite{Rubtsov12} and extended-DMFT 
\cite{PhysRevB.52.10295,Si1996,Smith00,Chitra01,Sun02} approaches, 
or from phonons.
A further route may be a proper design of the band structure, 
including multi-orbital models.
Another, completely different outcome of the frequency structure in the vertex is that it may possibly realize exotic gap functions on the frequency axis.

\begin{acknowledgments}
We thank P. Gunacker for providing quantum Monte Carlo results 
for comparison for the vertex and self-energy.  
We benefited from  
illuminating discussions with A. Toschi, G. Rohringer and P. Thunstr\"{o}m.
The present work was supported by 
European Research Council under the 
European Union's Seventh Framework Program
(FP/2007-2013) through ERC Grant No. 306447,
SFB ViCoM (M.K.,T.S.,K.H.),
Austrian Science Fund (FWF) through the 
Doctoral School ``Building Solids for Function" (T.S.)
and the Erwin-Schr\"odinger Fellowship J 4266 (SuMo, T.S.),
as well as by 
JSPS KAKENHI
Grant No. JP26247057 and ImPACT Program of Council 
for Science, Technology and Innovation, Cabinet Office, 
Government of Japan (Grant No. 2015-PM12-05-01) (M.K.,H.A.).  
T.S. also acknowledges 
the European Research Council
for the European Union Seventh Framework Program
(FP7/2007-2013) with ERC Grant No. 319286 (QMAC).
H.A. wishes to thank the Department of Physics, ETH 
Z\"{u}rich, Switzerland, for hospitality.
Calculations have been done mainly
on the Vienna Scientific Cluster (VSC).
\end{acknowledgments}

\bibliography{main,addrefs}

\begin{thebibliography}{67}%
\makeatletter
\providecommand \@ifxundefined [1]{%
 \@ifx{#1\undefined}
}%
\providecommand \@ifnum [1]{%
 \ifnum #1\expandafter \@firstoftwo
 \else \expandafter \@secondoftwo
 \fi
}%
\providecommand \@ifx [1]{%
 \ifx #1\expandafter \@firstoftwo
 \else \expandafter \@secondoftwo
 \fi
}%
\providecommand \natexlab [1]{#1}%
\providecommand \enquote  [1]{``#1''}%
\providecommand \bibnamefont  [1]{#1}%
\providecommand \bibfnamefont [1]{#1}%
\providecommand \citenamefont [1]{#1}%
\providecommand \href@noop [0]{\@secondoftwo}%
\providecommand \href [0]{\begingroup \@sanitize@url \@href}%
\providecommand \@href[1]{\@@startlink{#1}\@@href}%
\providecommand \@@href[1]{\endgroup#1\@@endlink}%
\providecommand \@sanitize@url [0]{\catcode `\\12\catcode `\$12\catcode
  `\&12\catcode `\#12\catcode `\^12\catcode `\_12\catcode `\%12\relax}%
\providecommand \@@startlink[1]{}%
\providecommand \@@endlink[0]{}%
\providecommand \url  [0]{\begingroup\@sanitize@url \@url }%
\providecommand \@url [1]{\endgroup\@href {#1}{\urlprefix }}%
\providecommand \urlprefix  [0]{URL }%
\providecommand \Eprint [0]{\href }%
\providecommand \doibase [0]{http://dx.doi.org/}%
\providecommand \selectlanguage [0]{\@gobble}%
\providecommand \bibinfo  [0]{\@secondoftwo}%
\providecommand \bibfield  [0]{\@secondoftwo}%
\providecommand \translation [1]{[#1]}%
\providecommand \BibitemOpen [0]{}%
\providecommand \bibitemStop [0]{}%
\providecommand \bibitemNoStop [0]{.\EOS\space}%
\providecommand \EOS [0]{\spacefactor3000\relax}%
\providecommand \BibitemShut  [1]{\csname bibitem#1\endcsname}%
\let\auto@bib@innerbib\@empty
%</preamble>
\bibitem [{\citenamefont {Bednorz}\ and\ \citenamefont
  {M\"uller}(1986)}]{Bednorz1986}%
  \BibitemOpen
  \bibfield  {author} {\bibinfo {author} {\bibfnamefont {J.~G.}\ \bibnamefont
  {Bednorz}}\ and\ \bibinfo {author} {\bibfnamefont {K.~A.}\ \bibnamefont
  {M\"uller}},\ }\href@noop {} {\bibfield  {journal} {\bibinfo  {journal}
  {Zeitschrift f\"ur Physik B Condensed Matter}\ }\textbf {\bibinfo {volume}
  {64}},\ \bibinfo {pages} {189} (\bibinfo {year} {1986})}\BibitemShut
  {NoStop}%
\bibitem [{\citenamefont {Schilling}\ \emph {et~al.}(1993)\citenamefont
  {Schilling}, \citenamefont {Cantoni}, \citenamefont {Guo},\ and\
  \citenamefont {Ott}}]{Schilling1993}%
  \BibitemOpen
  \bibfield  {author} {\bibinfo {author} {\bibfnamefont {A.}~\bibnamefont
  {Schilling}}, \bibinfo {author} {\bibfnamefont {M.}~\bibnamefont {Cantoni}},
  \bibinfo {author} {\bibfnamefont {J.~D.}\ \bibnamefont {Guo}}, \ and\
  \bibinfo {author} {\bibfnamefont {H.~R.}\ \bibnamefont {Ott}},\ }\href
  {\doibase 10.1038/363056a0} {\bibfield  {journal} {\bibinfo  {journal}
  {Nature}\ }\textbf {\bibinfo {volume} {363}},\ \bibinfo {pages} {56}
  (\bibinfo {year} {1993})}\BibitemShut {NoStop}%
\bibitem [{\citenamefont {Drozdov}\ \emph {et~al.}(2015)\citenamefont
  {Drozdov}, \citenamefont {Eremets}, \citenamefont {Troyan}, \citenamefont
  {Ksenofontov},\ and\ \citenamefont {Shylin}}]{Drozdov2015}%
  \BibitemOpen
  \bibfield  {author} {\bibinfo {author} {\bibfnamefont {A.~P.}\ \bibnamefont
  {Drozdov}}, \bibinfo {author} {\bibfnamefont {M.~I.}\ \bibnamefont
  {Eremets}}, \bibinfo {author} {\bibfnamefont {I.~A.}\ \bibnamefont {Troyan}},
  \bibinfo {author} {\bibfnamefont {V.}~\bibnamefont {Ksenofontov}}, \ and\
  \bibinfo {author} {\bibfnamefont {S.~I.}\ \bibnamefont {Shylin}},\ }\href
  {\doibase 10.1038/nature14964} {\bibfield  {journal} {\bibinfo  {journal}
  {Nature}\ }\textbf {\bibinfo {volume} {525}},\ \bibinfo {pages} {73}
  (\bibinfo {year} {2015})}\BibitemShut {NoStop}%
\bibitem [{\citenamefont {Scalapino}(2012)}]{Scalapino12}%
  \BibitemOpen
  \bibfield  {author} {\bibinfo {author} {\bibfnamefont {D.~J.}\ \bibnamefont
  {Scalapino}},\ }\href {\doibase 10.1103/RevModPhys.84.1383} {\bibfield
  {journal} {\bibinfo  {journal} {Rev. Mod. Phys.}\ }\textbf {\bibinfo {volume}
  {84}},\ \bibinfo {pages} {1383} (\bibinfo {year} {2012})}\BibitemShut
  {NoStop}%
\bibitem [{\citenamefont {Lee}\ and\ \citenamefont {Read}(1987)}]{Lee}%
  \BibitemOpen
  \bibfield  {author} {\bibinfo {author} {\bibfnamefont {P.~A.}\ \bibnamefont
  {Lee}}\ and\ \bibinfo {author} {\bibfnamefont {N.}~\bibnamefont {Read}},\
  }\href {\doibase 10.1103/PhysRevLett.58.2691} {\bibfield  {journal} {\bibinfo
   {journal} {Phys. Rev. Lett.}\ }\textbf {\bibinfo {volume} {58}},\ \bibinfo
  {pages} {2691} (\bibinfo {year} {1987})}\BibitemShut {NoStop}%
\bibitem [{\citenamefont {Kyung}\ \emph {et~al.}(2003)\citenamefont {Kyung},
  \citenamefont {Landry},\ and\ \citenamefont {Tremblay}}]{Kyung2003}%
  \BibitemOpen
  \bibfield  {author} {\bibinfo {author} {\bibfnamefont {B.}~\bibnamefont
  {Kyung}}, \bibinfo {author} {\bibfnamefont {J.-S.}\ \bibnamefont {Landry}}, \
  and\ \bibinfo {author} {\bibfnamefont {A.-M.~S.}\ \bibnamefont {Tremblay}},\
  }\href {\doibase 10.1103/PhysRevB.68.174502} {\bibfield  {journal} {\bibinfo
  {journal} {Phys. Rev. B}\ }\textbf {\bibinfo {volume} {68}},\ \bibinfo
  {pages} {174502} (\bibinfo {year} {2003})}\BibitemShut {NoStop}%
\bibitem [{\citenamefont {Maier}\ \emph {et~al.}(2005)\citenamefont {Maier},
  \citenamefont {Jarrell}, \citenamefont {Pruschke},\ and\ \citenamefont
  {Hettler}}]{Maier2005}%
  \BibitemOpen
  \bibfield  {author} {\bibinfo {author} {\bibfnamefont {T.}~\bibnamefont
  {Maier}}, \bibinfo {author} {\bibfnamefont {M.}~\bibnamefont {Jarrell}},
  \bibinfo {author} {\bibfnamefont {T.}~\bibnamefont {Pruschke}}, \ and\
  \bibinfo {author} {\bibfnamefont {M.~H.}\ \bibnamefont {Hettler}},\ }\href
  {\doibase 10.1103/RevModPhys.77.1027} {\bibfield  {journal} {\bibinfo
  {journal} {Rev. Mod. Phys.}\ }\textbf {\bibinfo {volume} {77}},\ \bibinfo
  {pages} {1027} (\bibinfo {year} {2005})}\BibitemShut {NoStop}%
\bibitem [{\citenamefont {Sakai}\ \emph {et~al.}(2016)\citenamefont {Sakai},
  \citenamefont {Civelli},\ and\ \citenamefont
  {Imada}}]{PhysRevLett.116.057003}%
  \BibitemOpen
  \bibfield  {author} {\bibinfo {author} {\bibfnamefont {S.}~\bibnamefont
  {Sakai}}, \bibinfo {author} {\bibfnamefont {M.}~\bibnamefont {Civelli}}, \
  and\ \bibinfo {author} {\bibfnamefont {M.}~\bibnamefont {Imada}},\ }\href
  {\doibase 10.1103/PhysRevLett.116.057003} {\bibfield  {journal} {\bibinfo
  {journal} {Phys. Rev. Lett.}\ }\textbf {\bibinfo {volume} {116}},\ \bibinfo
  {pages} {057003} (\bibinfo {year} {2016})}\BibitemShut {NoStop}%
\bibitem [{\citenamefont {Lee}\ \emph {et~al.}(2006)\citenamefont {Lee},
  \citenamefont {Nagaosa},\ and\ \citenamefont {Wen}}]{Lee2006}%
  \BibitemOpen
  \bibfield  {author} {\bibinfo {author} {\bibfnamefont {P.~A.}\ \bibnamefont
  {Lee}}, \bibinfo {author} {\bibfnamefont {N.}~\bibnamefont {Nagaosa}}, \ and\
  \bibinfo {author} {\bibfnamefont {X.-G.}\ \bibnamefont {Wen}},\ }\href
  {\doibase 10.1103/RevModPhys.78.17} {\bibfield  {journal} {\bibinfo
  {journal} {Rev. Mod. Phys.}\ }\textbf {\bibinfo {volume} {78}},\ \bibinfo
  {pages} {17} (\bibinfo {year} {2006})}\BibitemShut {NoStop}%
\bibitem [{\citenamefont {Hafermann}\ \emph {et~al.}(2009)\citenamefont
  {Hafermann}, \citenamefont {Kecker}, \citenamefont {Brener}, \citenamefont
  {Rubtsov}, \citenamefont {Katsnelson},\ and\ \citenamefont
  {Lichtenstein}}]{Hafermann2009a}%
  \BibitemOpen
  \bibfield  {author} {\bibinfo {author} {\bibfnamefont {H.}~\bibnamefont
  {Hafermann}}, \bibinfo {author} {\bibfnamefont {M.}~\bibnamefont {Kecker}},
  \bibinfo {author} {\bibfnamefont {S.}~\bibnamefont {Brener}}, \bibinfo
  {author} {\bibfnamefont {A.~N.}\ \bibnamefont {Rubtsov}}, \bibinfo {author}
  {\bibfnamefont {M.~I.}\ \bibnamefont {Katsnelson}}, \ and\ \bibinfo {author}
  {\bibfnamefont {A.~I.}\ \bibnamefont {Lichtenstein}},\ }\href
  {http://dx.doi.org/10.1007/s10948-008-0361-9} {\bibfield  {journal} {\bibinfo
   {journal} {J Supercond Nov Magn}\ }\textbf {\bibinfo {volume} {22}},\
  \bibinfo {pages} {45} (\bibinfo {year} {2009})}\BibitemShut {NoStop}%
\bibitem [{\citenamefont {Metzner}\ \emph {et~al.}(2012)\citenamefont
  {Metzner}, \citenamefont {Salmhofer}, \citenamefont {Honerkamp},
  \citenamefont {Meden},\ and\ \citenamefont {Sch\"onhammer}}]{Metzner2012}%
  \BibitemOpen
  \bibfield  {author} {\bibinfo {author} {\bibfnamefont {W.}~\bibnamefont
  {Metzner}}, \bibinfo {author} {\bibfnamefont {M.}~\bibnamefont {Salmhofer}},
  \bibinfo {author} {\bibfnamefont {C.}~\bibnamefont {Honerkamp}}, \bibinfo
  {author} {\bibfnamefont {V.}~\bibnamefont {Meden}}, \ and\ \bibinfo {author}
  {\bibfnamefont {K.}~\bibnamefont {Sch\"onhammer}},\ }\href {\doibase
  10.1103/RevModPhys.84.299} {\bibfield  {journal} {\bibinfo  {journal} {Rev.
  Mod. Phys.}\ }\textbf {\bibinfo {volume} {84}},\ \bibinfo {pages} {299}
  (\bibinfo {year} {2012})}\BibitemShut {NoStop}%
\bibitem [{\citenamefont {Otsuki}\ \emph {et~al.}(2014)\citenamefont {Otsuki},
  \citenamefont {Hafermann},\ and\ \citenamefont {Lichtenstein}}]{Otsuki2014}%
  \BibitemOpen
  \bibfield  {author} {\bibinfo {author} {\bibfnamefont {J.}~\bibnamefont
  {Otsuki}}, \bibinfo {author} {\bibfnamefont {H.}~\bibnamefont {Hafermann}}, \
  and\ \bibinfo {author} {\bibfnamefont {A.~I.}\ \bibnamefont {Lichtenstein}},\
  }\href {\doibase 10.1103/PhysRevB.90.235132} {\bibfield  {journal} {\bibinfo
  {journal} {Phys. Rev. B}\ }\textbf {\bibinfo {volume} {90}},\ \bibinfo
  {pages} {235132} (\bibinfo {year} {2014})}\BibitemShut {NoStop}%
\bibitem [{\citenamefont {Kitatani}\ \emph {et~al.}(2015)\citenamefont
  {Kitatani}, \citenamefont {Tsuji},\ and\ \citenamefont
  {Aoki}}]{Kitatani2015}%
  \BibitemOpen
  \bibfield  {author} {\bibinfo {author} {\bibfnamefont {M.}~\bibnamefont
  {Kitatani}}, \bibinfo {author} {\bibfnamefont {N.}~\bibnamefont {Tsuji}}, \
  and\ \bibinfo {author} {\bibfnamefont {H.}~\bibnamefont {Aoki}},\ }\href
  {http://link.aps.org/doi/10.1103/PhysRevB.92.085104} {\bibfield  {journal}
  {\bibinfo  {journal} {Phys. Rev. B}\ }\textbf {\bibinfo {volume} {92}},\
  \bibinfo {pages} {085104} (\bibinfo {year} {2015})}\BibitemShut {NoStop}%
\bibitem [{\citenamefont {Vu\ifmmode \check{c}\else \v{c}\fi{}i\ifmmode
  \check{c}\else \v{c}\fi{}evi\ifmmode~\acute{c}\else \'{c}\fi{}}\ \emph
  {et~al.}(2017)\citenamefont {Vu\ifmmode \check{c}\else \v{c}\fi{}i\ifmmode
  \check{c}\else \v{c}\fi{}evi\ifmmode~\acute{c}\else \'{c}\fi{}},
  \citenamefont {Ayral},\ and\ \citenamefont {Parcollet}}]{Vucicevic2017}%
  \BibitemOpen
  \bibfield  {author} {\bibinfo {author} {\bibfnamefont {J.}~\bibnamefont
  {Vu\ifmmode \check{c}\else \v{c}\fi{}i\ifmmode \check{c}\else
  \v{c}\fi{}evi\ifmmode~\acute{c}\else \'{c}\fi{}}}, \bibinfo {author}
  {\bibfnamefont {T.}~\bibnamefont {Ayral}}, \ and\ \bibinfo {author}
  {\bibfnamefont {O.}~\bibnamefont {Parcollet}},\ }\href {\doibase
  10.1103/PhysRevB.96.104504} {\bibfield  {journal} {\bibinfo  {journal} {Phys.
  Rev. B}\ }\textbf {\bibinfo {volume} {96}},\ \bibinfo {pages} {104504}
  (\bibinfo {year} {2017})}\BibitemShut {NoStop}%
\bibitem [{\citenamefont {L\"uders}\ \emph {et~al.}(2005)\citenamefont
  {L\"uders}, \citenamefont {Marques}, \citenamefont {Lathiotakis},
  \citenamefont {Floris}, \citenamefont {Profeta}, \citenamefont {Fast},
  \citenamefont {Continenza}, \citenamefont {Massidda},\ and\ \citenamefont
  {Gross}}]{SCDFT1}%
  \BibitemOpen
  \bibfield  {author} {\bibinfo {author} {\bibfnamefont {M.}~\bibnamefont
  {L\"uders}}, \bibinfo {author} {\bibfnamefont {M.~A.~L.}\ \bibnamefont
  {Marques}}, \bibinfo {author} {\bibfnamefont {N.~N.}\ \bibnamefont
  {Lathiotakis}}, \bibinfo {author} {\bibfnamefont {A.}~\bibnamefont {Floris}},
  \bibinfo {author} {\bibfnamefont {G.}~\bibnamefont {Profeta}}, \bibinfo
  {author} {\bibfnamefont {L.}~\bibnamefont {Fast}}, \bibinfo {author}
  {\bibfnamefont {A.}~\bibnamefont {Continenza}}, \bibinfo {author}
  {\bibfnamefont {S.}~\bibnamefont {Massidda}}, \ and\ \bibinfo {author}
  {\bibfnamefont {E.~K.~U.}\ \bibnamefont {Gross}},\ }\href {\doibase
  10.1103/PhysRevB.72.024545} {\bibfield  {journal} {\bibinfo  {journal} {Phys.
  Rev. B}\ }\textbf {\bibinfo {volume} {72}},\ \bibinfo {pages} {024545}
  (\bibinfo {year} {2005})}\BibitemShut {NoStop}%
\bibitem [{\citenamefont {Marques}\ \emph {et~al.}(2005)\citenamefont
  {Marques}, \citenamefont {L\"uders}, \citenamefont {Lathiotakis},
  \citenamefont {Profeta}, \citenamefont {Floris}, \citenamefont {Fast},
  \citenamefont {Continenza}, \citenamefont {Gross},\ and\ \citenamefont
  {Massidda}}]{SCDFT2}%
  \BibitemOpen
  \bibfield  {author} {\bibinfo {author} {\bibfnamefont {M.~A.~L.}\
  \bibnamefont {Marques}}, \bibinfo {author} {\bibfnamefont {M.}~\bibnamefont
  {L\"uders}}, \bibinfo {author} {\bibfnamefont {N.~N.}\ \bibnamefont
  {Lathiotakis}}, \bibinfo {author} {\bibfnamefont {G.}~\bibnamefont
  {Profeta}}, \bibinfo {author} {\bibfnamefont {A.}~\bibnamefont {Floris}},
  \bibinfo {author} {\bibfnamefont {L.}~\bibnamefont {Fast}}, \bibinfo {author}
  {\bibfnamefont {A.}~\bibnamefont {Continenza}}, \bibinfo {author}
  {\bibfnamefont {E.~K.~U.}\ \bibnamefont {Gross}}, \ and\ \bibinfo {author}
  {\bibfnamefont {S.}~\bibnamefont {Massidda}},\ }\href {\doibase
  10.1103/PhysRevB.72.024546} {\bibfield  {journal} {\bibinfo  {journal} {Phys.
  Rev. B}\ }\textbf {\bibinfo {volume} {72}},\ \bibinfo {pages} {024546}
  (\bibinfo {year} {2005})}\BibitemShut {NoStop}%
\bibitem [{\citenamefont {Migdal}(1958)}]{Migdal1958}%
  \BibitemOpen
  \bibfield  {author} {\bibinfo {author} {\bibfnamefont {A.~B.}\ \bibnamefont
  {Migdal}},\ }\href {http://www.jetp.ac.ru/cgi-bin/e/index/e/7/6/p996?a=list}
  {\bibfield  {journal} {\bibinfo  {journal} {Sov. Phys. JETP}\ }\textbf
  {\bibinfo {volume} {7}},\ \bibinfo {pages} {996} (\bibinfo {year}
  {1958})}\BibitemShut {NoStop}%
\bibitem [{\citenamefont {Onari}\ \emph {et~al.}(2014)\citenamefont {Onari},
  \citenamefont {Yamakawa},\ and\ \citenamefont
  {Kontani}}]{PhysRevLett.112.187001}%
  \BibitemOpen
  \bibfield  {author} {\bibinfo {author} {\bibfnamefont {S.}~\bibnamefont
  {Onari}}, \bibinfo {author} {\bibfnamefont {Y.}~\bibnamefont {Yamakawa}}, \
  and\ \bibinfo {author} {\bibfnamefont {H.}~\bibnamefont {Kontani}},\ }\href
  {\doibase 10.1103/PhysRevLett.112.187001} {\bibfield  {journal} {\bibinfo
  {journal} {Phys. Rev. Lett.}\ }\textbf {\bibinfo {volume} {112}},\ \bibinfo
  {pages} {187001} (\bibinfo {year} {2014})}\BibitemShut {NoStop}%
\bibitem [{\citenamefont {Vilardi}\ \emph {et~al.}(2017)\citenamefont
  {Vilardi}, \citenamefont {Taranto},\ and\ \citenamefont
  {Metzner}}]{PhysRevB.96.235110}%
  \BibitemOpen
  \bibfield  {author} {\bibinfo {author} {\bibfnamefont {D.}~\bibnamefont
  {Vilardi}}, \bibinfo {author} {\bibfnamefont {C.}~\bibnamefont {Taranto}}, \
  and\ \bibinfo {author} {\bibfnamefont {W.}~\bibnamefont {Metzner}},\ }\href
  {\doibase 10.1103/PhysRevB.96.235110} {\bibfield  {journal} {\bibinfo
  {journal} {Phys. Rev. B}\ }\textbf {\bibinfo {volume} {96}},\ \bibinfo
  {pages} {235110} (\bibinfo {year} {2017})}\BibitemShut {NoStop}%
\bibitem [{\citenamefont {Metzner}\ and\ \citenamefont
  {Vollhardt}(1989)}]{Metzner1989}%
  \BibitemOpen
  \bibfield  {author} {\bibinfo {author} {\bibfnamefont {W.}~\bibnamefont
  {Metzner}}\ and\ \bibinfo {author} {\bibfnamefont {D.}~\bibnamefont
  {Vollhardt}},\ }\href {\doibase 10.1103/PhysRevLett.62.324} {\bibfield
  {journal} {\bibinfo  {journal} {Phys. Rev. Lett.}\ }\textbf {\bibinfo
  {volume} {62}},\ \bibinfo {pages} {324} (\bibinfo {year} {1989})}\BibitemShut
  {NoStop}%
\bibitem [{\citenamefont {Georges}\ and\ \citenamefont
  {Kotliar}(1992)}]{Georges1992a}%
  \BibitemOpen
  \bibfield  {author} {\bibinfo {author} {\bibfnamefont {A.}~\bibnamefont
  {Georges}}\ and\ \bibinfo {author} {\bibfnamefont {G.}~\bibnamefont
  {Kotliar}},\ }\href {\doibase 10.1103/PhysRevB.45.6479} {\bibfield  {journal}
  {\bibinfo  {journal} {Phys. Rev. B}\ }\textbf {\bibinfo {volume} {45}},\
  \bibinfo {pages} {6479} (\bibinfo {year} {1992})}\BibitemShut {NoStop}%
\bibitem [{\citenamefont {Georges}\ \emph {et~al.}(1996)\citenamefont
  {Georges}, \citenamefont {Kotliar}, \citenamefont {Krauth},\ and\
  \citenamefont {Rozenberg}}]{Georges1996}%
  \BibitemOpen
  \bibfield  {author} {\bibinfo {author} {\bibfnamefont {A.}~\bibnamefont
  {Georges}}, \bibinfo {author} {\bibfnamefont {G.}~\bibnamefont {Kotliar}},
  \bibinfo {author} {\bibfnamefont {W.}~\bibnamefont {Krauth}}, \ and\ \bibinfo
  {author} {\bibfnamefont {M.~J.}\ \bibnamefont {Rozenberg}},\ }\href {\doibase
  10.1103/RevModPhys.68.13} {\bibfield  {journal} {\bibinfo  {journal} {Rev.
  Mod. Phys.}\ }\textbf {\bibinfo {volume} {68}},\ \bibinfo {pages} {13}
  (\bibinfo {year} {1996})}\BibitemShut {NoStop}%
\bibitem [{\citenamefont {Toschi}\ \emph {et~al.}(2007)\citenamefont {Toschi},
  \citenamefont {Katanin},\ and\ \citenamefont {Held}}]{Toschi2007}%
  \BibitemOpen
  \bibfield  {author} {\bibinfo {author} {\bibfnamefont {A.}~\bibnamefont
  {Toschi}}, \bibinfo {author} {\bibfnamefont {A.~A.}\ \bibnamefont {Katanin}},
  \ and\ \bibinfo {author} {\bibfnamefont {K.}~\bibnamefont {Held}},\ }\href
  {\doibase 10.1103/PhysRevB.75.045118} {\bibfield  {journal} {\bibinfo
  {journal} {Phys Rev. B}\ }\textbf {\bibinfo {volume} {75}},\ \bibinfo {pages}
  {045118} (\bibinfo {year} {2007})}\BibitemShut {NoStop}%
\bibitem [{\citenamefont {Kusunose}(2006)}]{Kusunose2006}%
  \BibitemOpen
  \bibfield  {author} {\bibinfo {author} {\bibfnamefont {H.}~\bibnamefont
  {Kusunose}},\ }\href {\doibase 10.1143/JPSJ.75.054713} {\bibfield  {journal}
  {\bibinfo  {journal} {J. Phys. Soc. Jpn.}\ }\textbf {\bibinfo {volume}
  {75}},\ \bibinfo {pages} {054713} (\bibinfo {year} {2006})}\BibitemShut
  {NoStop}%
\bibitem [{\citenamefont {Slezak}\ \emph {et~al.}(2009)\citenamefont {Slezak},
  \citenamefont {Jarrell}, \citenamefont {Maier},\ and\ \citenamefont
  {Deisz}}]{Slezak2009}%
  \BibitemOpen
  \bibfield  {author} {\bibinfo {author} {\bibfnamefont {C.}~\bibnamefont
  {Slezak}}, \bibinfo {author} {\bibfnamefont {M.}~\bibnamefont {Jarrell}},
  \bibinfo {author} {\bibfnamefont {T.}~\bibnamefont {Maier}}, \ and\ \bibinfo
  {author} {\bibfnamefont {J.}~\bibnamefont {Deisz}},\ }\href
  {http://stacks.iop.org/0953-8984/21/i=43/a=435604} {\bibfield  {journal}
  {\bibinfo  {journal} {J. Phys.: Condens. Matter}\ }\textbf {\bibinfo {volume}
  {21}},\ \bibinfo {pages} {435604} (\bibinfo {year} {2009})}\BibitemShut
  {NoStop}%
\bibitem [{\citenamefont {Katanin}\ \emph {et~al.}(2009)\citenamefont
  {Katanin}, \citenamefont {Toschi},\ and\ \citenamefont {Held}}]{Katanin2009}%
  \BibitemOpen
  \bibfield  {author} {\bibinfo {author} {\bibfnamefont {A.~A.}\ \bibnamefont
  {Katanin}}, \bibinfo {author} {\bibfnamefont {A.}~\bibnamefont {Toschi}}, \
  and\ \bibinfo {author} {\bibfnamefont {K.}~\bibnamefont {Held}},\ }\href
  {\doibase 10.1103/PhysRevB.80.075104} {\bibfield  {journal} {\bibinfo
  {journal} {Phys. Rev. B}\ }\textbf {\bibinfo {volume} {80}},\ \bibinfo
  {pages} {075104} (\bibinfo {year} {2009})}\BibitemShut {NoStop}%
\bibitem [{\citenamefont {Rubtsov}\ \emph {et~al.}(2008)\citenamefont
  {Rubtsov}, \citenamefont {Katsnelson},\ and\ \citenamefont
  {Lichtenstein}}]{Rubtsov2008}%
  \BibitemOpen
  \bibfield  {author} {\bibinfo {author} {\bibfnamefont {A.~N.}\ \bibnamefont
  {Rubtsov}}, \bibinfo {author} {\bibfnamefont {M.~I.}\ \bibnamefont
  {Katsnelson}}, \ and\ \bibinfo {author} {\bibfnamefont {A.~I.}\ \bibnamefont
  {Lichtenstein}},\ }\href {\doibase 10.1103/PhysRevB.77.033101} {\bibfield
  {journal} {\bibinfo  {journal} {Phys. Rev. B}\ }\textbf {\bibinfo {volume}
  {77}},\ \bibinfo {pages} {033101} (\bibinfo {year} {2008})}\BibitemShut
  {NoStop}%
\bibitem [{\citenamefont {Rohringer}\ \emph {et~al.}(2013)\citenamefont
  {Rohringer}, \citenamefont {Toschi}, \citenamefont {Hafermann}, \citenamefont
  {Held}, \citenamefont {Anisimov},\ and\ \citenamefont
  {Katanin}}]{Rohringer2013}%
  \BibitemOpen
  \bibfield  {author} {\bibinfo {author} {\bibfnamefont {G.}~\bibnamefont
  {Rohringer}}, \bibinfo {author} {\bibfnamefont {A.}~\bibnamefont {Toschi}},
  \bibinfo {author} {\bibfnamefont {H.}~\bibnamefont {Hafermann}}, \bibinfo
  {author} {\bibfnamefont {K.}~\bibnamefont {Held}}, \bibinfo {author}
  {\bibfnamefont {V.~I.}\ \bibnamefont {Anisimov}}, \ and\ \bibinfo {author}
  {\bibfnamefont {A.~A.}\ \bibnamefont {Katanin}},\ }\href
  {http://link.aps.org/doi/10.1103/PhysRevB.88.115112} {\bibfield  {journal}
  {\bibinfo  {journal} {Phys. Rev. B}\ }\textbf {\bibinfo {volume} {88}},\
  \bibinfo {pages} {115112} (\bibinfo {year} {2013})}\BibitemShut {NoStop}%
\bibitem [{\citenamefont {Taranto}\ \emph {et~al.}(2014)\citenamefont
  {Taranto}, \citenamefont {Andergassen}, \citenamefont {Bauer}, \citenamefont
  {Held}, \citenamefont {Katanin}, \citenamefont {Metzner}, \citenamefont
  {Rohringer},\ and\ \citenamefont {Toschi}}]{Taranto2014}%
  \BibitemOpen
  \bibfield  {author} {\bibinfo {author} {\bibfnamefont {C.}~\bibnamefont
  {Taranto}}, \bibinfo {author} {\bibfnamefont {S.}~\bibnamefont
  {Andergassen}}, \bibinfo {author} {\bibfnamefont {J.}~\bibnamefont {Bauer}},
  \bibinfo {author} {\bibfnamefont {K.}~\bibnamefont {Held}}, \bibinfo {author}
  {\bibfnamefont {A.}~\bibnamefont {Katanin}}, \bibinfo {author} {\bibfnamefont
  {W.}~\bibnamefont {Metzner}}, \bibinfo {author} {\bibfnamefont
  {G.}~\bibnamefont {Rohringer}}, \ and\ \bibinfo {author} {\bibfnamefont
  {A.}~\bibnamefont {Toschi}},\ }\href {\doibase
  10.1103/PhysRevLett.112.196402} {\bibfield  {journal} {\bibinfo  {journal}
  {Phys. Rev. Lett.}\ }\textbf {\bibinfo {volume} {112}},\ \bibinfo {pages}
  {196402} (\bibinfo {year} {2014})}\BibitemShut {NoStop}%
\bibitem [{\citenamefont {Ayral}\ and\ \citenamefont
  {Parcollet}(2015)}]{Ayral2015}%
  \BibitemOpen
  \bibfield  {author} {\bibinfo {author} {\bibfnamefont {T.}~\bibnamefont
  {Ayral}}\ and\ \bibinfo {author} {\bibfnamefont {O.}~\bibnamefont
  {Parcollet}},\ }\href {http://link.aps.org/doi/10.1103/PhysRevB.92.115109}
  {\bibfield  {journal} {\bibinfo  {journal} {Phys Rev. B}\ }\textbf {\bibinfo
  {volume} {92}},\ \bibinfo {pages} {115109} (\bibinfo {year}
  {2015})}\BibitemShut {NoStop}%
\bibitem [{\citenamefont {Li}(2015)}]{Li2015}%
  \BibitemOpen
  \bibfield  {author} {\bibinfo {author} {\bibfnamefont {G.}~\bibnamefont
  {Li}},\ }\href {\doibase 10.1103/PhysRevB.91.165134} {\bibfield  {journal}
  {\bibinfo  {journal} {Phys. Rev. B}\ }\textbf {\bibinfo {volume} {91}},\
  \bibinfo {pages} {165134} (\bibinfo {year} {2015})}\BibitemShut {NoStop}%
\bibitem [{\citenamefont {Ayral}\ and\ \citenamefont
  {Parcollet}(2016{\natexlab{a}})}]{Ayral2016a}%
  \BibitemOpen
  \bibfield  {author} {\bibinfo {author} {\bibfnamefont {T.}~\bibnamefont
  {Ayral}}\ and\ \bibinfo {author} {\bibfnamefont {O.}~\bibnamefont
  {Parcollet}},\ }\href {\doibase 10.1103/PhysRevB.93.235124} {\bibfield
  {journal} {\bibinfo  {journal} {Phys. Rev. B}\ }\textbf {\bibinfo {volume}
  {93}},\ \bibinfo {pages} {235124} (\bibinfo {year}
  {2016}{\natexlab{a}})}\BibitemShut {NoStop}%
\bibitem [{\citenamefont {Ayral}\ and\ \citenamefont
  {Parcollet}(2016{\natexlab{b}})}]{Ayral2016}%
  \BibitemOpen
  \bibfield  {author} {\bibinfo {author} {\bibfnamefont {T.}~\bibnamefont
  {Ayral}}\ and\ \bibinfo {author} {\bibfnamefont {O.}~\bibnamefont
  {Parcollet}},\ }\href {\doibase 10.1103/PhysRevB.94.075159} {\bibfield
  {journal} {\bibinfo  {journal} {Phys. Rev. B}\ }\textbf {\bibinfo {volume}
  {94}},\ \bibinfo {pages} {075159} (\bibinfo {year}
  {2016}{\natexlab{b}})}\BibitemShut {NoStop}%
\bibitem [{\citenamefont {Rohringer}\ \emph {et~al.}(2018)\citenamefont
  {Rohringer}, \citenamefont {Hafermann}, \citenamefont {Toschi}, \citenamefont
  {Katanin}, \citenamefont {Antipov}, \citenamefont {Katsnelson}, \citenamefont
  {Lichtenstein}, \citenamefont {Rubtsov},\ and\ \citenamefont
  {Held}}]{RMPVertex}%
  \BibitemOpen
  \bibfield  {author} {\bibinfo {author} {\bibfnamefont {G.}~\bibnamefont
  {Rohringer}}, \bibinfo {author} {\bibfnamefont {H.}~\bibnamefont
  {Hafermann}}, \bibinfo {author} {\bibfnamefont {A.}~\bibnamefont {Toschi}},
  \bibinfo {author} {\bibfnamefont {A.~A.}\ \bibnamefont {Katanin}}, \bibinfo
  {author} {\bibfnamefont {A.~E.}\ \bibnamefont {Antipov}}, \bibinfo {author}
  {\bibfnamefont {M.~I.}\ \bibnamefont {Katsnelson}}, \bibinfo {author}
  {\bibfnamefont {A.~I.}\ \bibnamefont {Lichtenstein}}, \bibinfo {author}
  {\bibfnamefont {A.~N.}\ \bibnamefont {Rubtsov}}, \ and\ \bibinfo {author}
  {\bibfnamefont {K.}~\bibnamefont {Held}},\ }\href {\doibase
  10.1103/RevModPhys.90.025003} {\bibfield  {journal} {\bibinfo  {journal}
  {Rev. Mod. Phys.}\ }\textbf {\bibinfo {volume} {90}},\ \bibinfo {pages}
  {025003} (\bibinfo {year} {2018})}\BibitemShut {NoStop}%
\bibitem [{\citenamefont {Rohringer}\ \emph {et~al.}(2012)\citenamefont
  {Rohringer}, \citenamefont {Valli},\ and\ \citenamefont
  {Toschi}}]{Rohringer2012}%
  \BibitemOpen
  \bibfield  {author} {\bibinfo {author} {\bibfnamefont {G.}~\bibnamefont
  {Rohringer}}, \bibinfo {author} {\bibfnamefont {A.}~\bibnamefont {Valli}}, \
  and\ \bibinfo {author} {\bibfnamefont {A.}~\bibnamefont {Toschi}},\ }\href
  {\doibase 10.1103/PhysRevB.86.125114} {\bibfield  {journal} {\bibinfo
  {journal} {Phys. Rev. B}\ }\textbf {\bibinfo {volume} {86}},\ \bibinfo
  {pages} {125114} (\bibinfo {year} {2012})}\BibitemShut {NoStop}%
\bibitem [{\citenamefont {Rohringer}(2013)}]{Rohringer2013a}%
  \BibitemOpen
  \bibfield  {author} {\bibinfo {author} {\bibfnamefont {G.}~\bibnamefont
  {Rohringer}},\ }\emph {\bibinfo {title} {New routes towards a theoretical
  treatment of nonlocal electronic correlations}},\ \href@noop {} {Ph.D.
  thesis},\ \bibinfo  {school} {Vienna University of Technology} (\bibinfo
  {year} {2013})\BibitemShut {NoStop}%
\bibitem [{\citenamefont {Wentzell}\ \emph {et~al.}(2016)\citenamefont
  {Wentzell}, \citenamefont {Li}, \citenamefont {Tagliavini}, \citenamefont
  {Taranto}, \citenamefont {Rohringer}, \citenamefont {Held}, \citenamefont
  {Toschi},\ and\ \citenamefont {Andergassen}}]{Wentzell2016}%
  \BibitemOpen
  \bibfield  {author} {\bibinfo {author} {\bibfnamefont {N.}~\bibnamefont
  {Wentzell}}, \bibinfo {author} {\bibfnamefont {G.}~\bibnamefont {Li}},
  \bibinfo {author} {\bibfnamefont {A.}~\bibnamefont {Tagliavini}}, \bibinfo
  {author} {\bibfnamefont {C.}~\bibnamefont {Taranto}}, \bibinfo {author}
  {\bibfnamefont {G.}~\bibnamefont {Rohringer}}, \bibinfo {author}
  {\bibfnamefont {K.}~\bibnamefont {Held}}, \bibinfo {author} {\bibfnamefont
  {A.}~\bibnamefont {Toschi}}, \ and\ \bibinfo {author} {\bibfnamefont
  {S.}~\bibnamefont {Andergassen}},\ }\href@noop {} {\bibfield  {journal}
  {\bibinfo  {journal} {arXiv:1610.06520}\ } (\bibinfo {year}
  {2016})}\BibitemShut {NoStop}%
\bibitem [{\citenamefont {Rubtsov}\ \emph {et~al.}(2009)\citenamefont
  {Rubtsov}, \citenamefont {Katsnelson}, \citenamefont {Lichtenstein},\ and\
  \citenamefont {Georges}}]{Rubtsov2009}%
  \BibitemOpen
  \bibfield  {author} {\bibinfo {author} {\bibfnamefont {A.~N.}\ \bibnamefont
  {Rubtsov}}, \bibinfo {author} {\bibfnamefont {M.~I.}\ \bibnamefont
  {Katsnelson}}, \bibinfo {author} {\bibfnamefont {A.~I.}\ \bibnamefont
  {Lichtenstein}}, \ and\ \bibinfo {author} {\bibfnamefont {A.}~\bibnamefont
  {Georges}},\ }\href {\doibase 10.1103/PhysRevB.79.045133} {\bibfield
  {journal} {\bibinfo  {journal} {Phys. Rev. B}\ }\textbf {\bibinfo {volume}
  {79}},\ \bibinfo {pages} {045133} (\bibinfo {year} {2009})}\BibitemShut
  {NoStop}%
\bibitem [{\citenamefont {Sch\"afer}\ \emph {et~al.}(2015)\citenamefont
  {Sch\"afer}, \citenamefont {Geles}, \citenamefont {Rost}, \citenamefont
  {Rohringer}, \citenamefont {Arrigoni}, \citenamefont {Held}, \citenamefont
  {Bl\"umer}, \citenamefont {Aichhorn},\ and\ \citenamefont
  {Toschi}}]{Schaefer2015-2}%
  \BibitemOpen
  \bibfield  {author} {\bibinfo {author} {\bibfnamefont {T.}~\bibnamefont
  {Sch\"afer}}, \bibinfo {author} {\bibfnamefont {F.}~\bibnamefont {Geles}},
  \bibinfo {author} {\bibfnamefont {D.}~\bibnamefont {Rost}}, \bibinfo {author}
  {\bibfnamefont {G.}~\bibnamefont {Rohringer}}, \bibinfo {author}
  {\bibfnamefont {E.}~\bibnamefont {Arrigoni}}, \bibinfo {author}
  {\bibfnamefont {K.}~\bibnamefont {Held}}, \bibinfo {author} {\bibfnamefont
  {N.}~\bibnamefont {Bl\"umer}}, \bibinfo {author} {\bibfnamefont
  {M.}~\bibnamefont {Aichhorn}}, \ and\ \bibinfo {author} {\bibfnamefont
  {A.}~\bibnamefont {Toschi}},\ }\href {\doibase 10.1103/PhysRevB.91.125109}
  {\bibfield  {journal} {\bibinfo  {journal} {Phys. Rev. B}\ }\textbf {\bibinfo
  {volume} {91}},\ \bibinfo {pages} {125109} (\bibinfo {year}
  {2015})}\BibitemShut {NoStop}%
\bibitem [{\citenamefont {Sch\"afer}\ \emph {et~al.}(2016)\citenamefont
  {Sch\"afer}, \citenamefont {Toschi},\ and\ \citenamefont
  {Held}}]{Schaefer2015-3}%
  \BibitemOpen
  \bibfield  {author} {\bibinfo {author} {\bibfnamefont {T.}~\bibnamefont
  {Sch\"afer}}, \bibinfo {author} {\bibfnamefont {A.}~\bibnamefont {Toschi}}, \
  and\ \bibinfo {author} {\bibfnamefont {K.}~\bibnamefont {Held}},\ }\href
  {\doibase 10.1016/j.jmmm.2015.07.103} {\bibfield  {journal} {\bibinfo
  {journal} {Journal of Magnetism and Magnetic Materials}\ }\textbf {\bibinfo
  {volume} {400}},\ \bibinfo {pages} {107} (\bibinfo {year}
  {2016})}\BibitemShut {NoStop}%
\bibitem [{\citenamefont {Gunnarsson}\ \emph {et~al.}(2015)\citenamefont
  {Gunnarsson}, \citenamefont {Sch\"afer}, \citenamefont {LeBlanc},
  \citenamefont {Gull}, \citenamefont {Merino}, \citenamefont {Sangiovanni},
  \citenamefont {Rohringer},\ and\ \citenamefont {Toschi}}]{Gunnarsson2015}%
  \BibitemOpen
  \bibfield  {author} {\bibinfo {author} {\bibfnamefont {O.}~\bibnamefont
  {Gunnarsson}}, \bibinfo {author} {\bibfnamefont {T.}~\bibnamefont
  {Sch\"afer}}, \bibinfo {author} {\bibfnamefont {J.~P.~F.}\ \bibnamefont
  {LeBlanc}}, \bibinfo {author} {\bibfnamefont {E.}~\bibnamefont {Gull}},
  \bibinfo {author} {\bibfnamefont {J.}~\bibnamefont {Merino}}, \bibinfo
  {author} {\bibfnamefont {G.}~\bibnamefont {Sangiovanni}}, \bibinfo {author}
  {\bibfnamefont {G.}~\bibnamefont {Rohringer}}, \ and\ \bibinfo {author}
  {\bibfnamefont {A.}~\bibnamefont {Toschi}},\ }\href {\doibase
  10.1103/PhysRevLett.114.236402} {\bibfield  {journal} {\bibinfo  {journal}
  {Phys. Rev. Lett.}\ }\textbf {\bibinfo {volume} {114}},\ \bibinfo {pages}
  {236402} (\bibinfo {year} {2015})}\BibitemShut {NoStop}%
\bibitem [{\citenamefont {Pudleiner}\ \emph {et~al.}(2016)\citenamefont
  {Pudleiner}, \citenamefont {Sch\"afer}, \citenamefont {Rost}, \citenamefont
  {Li}, \citenamefont {Held},\ and\ \citenamefont {Bl\"umer}}]{Pudleiner2016}%
  \BibitemOpen
  \bibfield  {author} {\bibinfo {author} {\bibfnamefont {P.}~\bibnamefont
  {Pudleiner}}, \bibinfo {author} {\bibfnamefont {T.}~\bibnamefont
  {Sch\"afer}}, \bibinfo {author} {\bibfnamefont {D.}~\bibnamefont {Rost}},
  \bibinfo {author} {\bibfnamefont {G.}~\bibnamefont {Li}}, \bibinfo {author}
  {\bibfnamefont {K.}~\bibnamefont {Held}}, \ and\ \bibinfo {author}
  {\bibfnamefont {N.}~\bibnamefont {Bl\"umer}},\ }\href {\doibase
  10.1103/PhysRevB.93.195134} {\bibfield  {journal} {\bibinfo  {journal} {Phys.
  Rev. B}\ }\textbf {\bibinfo {volume} {93}},\ \bibinfo {pages} {195134}
  (\bibinfo {year} {2016})}\BibitemShut {NoStop}%
\bibitem [{\citenamefont {Gunnarsson}\ \emph {et~al.}(2018)\citenamefont
  {Gunnarsson}, \citenamefont {Merino}, \citenamefont {Sch\"afer},
  \citenamefont {Sangiovanni}, \citenamefont {Rohringer},\ and\ \citenamefont
  {Toschi}}]{gunnarsson2017complementary}%
  \BibitemOpen
  \bibfield  {author} {\bibinfo {author} {\bibfnamefont {O.}~\bibnamefont
  {Gunnarsson}}, \bibinfo {author} {\bibfnamefont {J.}~\bibnamefont {Merino}},
  \bibinfo {author} {\bibfnamefont {T.}~\bibnamefont {Sch\"afer}}, \bibinfo
  {author} {\bibfnamefont {G.}~\bibnamefont {Sangiovanni}}, \bibinfo {author}
  {\bibfnamefont {G.}~\bibnamefont {Rohringer}}, \ and\ \bibinfo {author}
  {\bibfnamefont {A.}~\bibnamefont {Toschi}},\ }\href {\doibase
  10.1103/PhysRevB.97.125134} {\bibfield  {journal} {\bibinfo  {journal} {Phys.
  Rev. B}\ }\textbf {\bibinfo {volume} {97}},\ \bibinfo {pages} {125134}
  (\bibinfo {year} {2018})}\BibitemShut {NoStop}%
\bibitem [{\citenamefont {Nishiguchi}\ \emph {et~al.}(2013)\citenamefont
  {Nishiguchi}, \citenamefont {Kuroki}, \citenamefont {Arita}, \citenamefont
  {Oka},\ and\ \citenamefont {Aoki}}]{nishiguchi2013}%
  \BibitemOpen
  \bibfield  {author} {\bibinfo {author} {\bibfnamefont {K.}~\bibnamefont
  {Nishiguchi}}, \bibinfo {author} {\bibfnamefont {K.}~\bibnamefont {Kuroki}},
  \bibinfo {author} {\bibfnamefont {R.}~\bibnamefont {Arita}}, \bibinfo
  {author} {\bibfnamefont {T.}~\bibnamefont {Oka}}, \ and\ \bibinfo {author}
  {\bibfnamefont {H.}~\bibnamefont {Aoki}},\ }\href {\doibase
  10.1103/PhysRevB.88.014509} {\bibfield  {journal} {\bibinfo  {journal} {Phys.
  Rev. B}\ }\textbf {\bibinfo {volume} {88}},\ \bibinfo {pages} {014509}
  (\bibinfo {year} {2013})}\BibitemShut {NoStop}%
\bibitem [{\citenamefont {Nishiguchi}(2012)}]{nishiguchiphd}%
  \BibitemOpen
  \bibfield  {author} {\bibinfo {author} {\bibfnamefont {K.}~\bibnamefont
  {Nishiguchi}},\ }\emph {\bibinfo {title} {Theory of high $T_{\rm c}$
  superconductivity in multi-layered cuprates}},\ \href@noop {} {Ph.D.
  thesis},\ \bibinfo  {school} {University of Tokyo} (\bibinfo {year}
  {2012})\BibitemShut {NoStop}%
\bibitem [{\citenamefont {Werner}\ \emph {et~al.}(2006)\citenamefont {Werner},
  \citenamefont {Comanac}, \citenamefont {de' Medici}, \citenamefont {Troyer},\
  and\ \citenamefont {Millis}}]{CTHYB1}%
  \BibitemOpen
  \bibfield  {author} {\bibinfo {author} {\bibfnamefont {P.}~\bibnamefont
  {Werner}}, \bibinfo {author} {\bibfnamefont {A.}~\bibnamefont {Comanac}},
  \bibinfo {author} {\bibfnamefont {L.}~\bibnamefont {de' Medici}}, \bibinfo
  {author} {\bibfnamefont {M.}~\bibnamefont {Troyer}}, \ and\ \bibinfo {author}
  {\bibfnamefont {A.~J.}\ \bibnamefont {Millis}},\ }\href {\doibase
  10.1103/PhysRevLett.97.076405} {\bibfield  {journal} {\bibinfo  {journal}
  {Phys. Rev. Lett.}\ }\textbf {\bibinfo {volume} {97}},\ \bibinfo {pages}
  {076405} (\bibinfo {year} {2006})}\BibitemShut {NoStop}%
\bibitem [{\citenamefont {Werner}\ and\ \citenamefont {Millis}(2006)}]{CTHYB2}%
  \BibitemOpen
  \bibfield  {author} {\bibinfo {author} {\bibfnamefont {P.}~\bibnamefont
  {Werner}}\ and\ \bibinfo {author} {\bibfnamefont {A.~J.}\ \bibnamefont
  {Millis}},\ }\href {\doibase 10.1103/PhysRevB.74.155107} {\bibfield
  {journal} {\bibinfo  {journal} {Phys. Rev. B}\ }\textbf {\bibinfo {volume}
  {74}},\ \bibinfo {pages} {155107} (\bibinfo {year} {2006})}\BibitemShut
  {NoStop}%
\bibitem [{\citenamefont {Parragh}\ \emph {et~al.}(2012)\citenamefont
  {Parragh}, \citenamefont {Toschi}, \citenamefont {Held},\ and\ \citenamefont
  {Sangiovanni}}]{Parragh2012}%
  \BibitemOpen
  \bibfield  {author} {\bibinfo {author} {\bibfnamefont {N.}~\bibnamefont
  {Parragh}}, \bibinfo {author} {\bibfnamefont {A.}~\bibnamefont {Toschi}},
  \bibinfo {author} {\bibfnamefont {K.}~\bibnamefont {Held}}, \ and\ \bibinfo
  {author} {\bibfnamefont {G.}~\bibnamefont {Sangiovanni}},\ }\href {\doibase
  10.1103/PhysRevB.86.155158} {\bibfield  {journal} {\bibinfo  {journal} {Phys.
  Rev. B}\ }\textbf {\bibinfo {volume} {86}},\ \bibinfo {pages} {155158}
  (\bibinfo {year} {2012})}\BibitemShut {NoStop}%
\bibitem [{\citenamefont {Wallerberger}\ \emph {et~al.}(2019)\citenamefont
  {Wallerberger}, \citenamefont {Hausoel}, \citenamefont {Gunacker},
  \citenamefont {Kowalski}, \citenamefont {Parragh}, \citenamefont {Goth},
  \citenamefont {Held},\ and\ \citenamefont {Sangiovanni}}]{W2DYN}%
  \BibitemOpen
  \bibfield  {author} {\bibinfo {author} {\bibfnamefont {M.}~\bibnamefont
  {Wallerberger}}, \bibinfo {author} {\bibfnamefont {A.}~\bibnamefont
  {Hausoel}}, \bibinfo {author} {\bibfnamefont {P.}~\bibnamefont {Gunacker}},
  \bibinfo {author} {\bibfnamefont {A.}~\bibnamefont {Kowalski}}, \bibinfo
  {author} {\bibfnamefont {N.}~\bibnamefont {Parragh}}, \bibinfo {author}
  {\bibfnamefont {F.}~\bibnamefont {Goth}}, \bibinfo {author} {\bibfnamefont
  {K.}~\bibnamefont {Held}}, \ and\ \bibinfo {author} {\bibfnamefont
  {G.}~\bibnamefont {Sangiovanni}},\ }\href@noop {} {\bibfield  {journal}
  {\bibinfo  {journal} {Computer Physics Communications}\ }\textbf {\bibinfo
  {volume} {235}},\ \bibinfo {pages} {388 } (\bibinfo {year}
  {2019})}\BibitemShut {NoStop}%
\bibitem [{sup()}]{suppl}%
  \BibitemOpen
  \href@noop {} {}\bibinfo {note} {Section S.1 in the Supplemental Material
  provides further details on the relation between $\Gamma_m$ and $F$,
  including the $\lambda$-correction of the physical susceptibility. In Section
  S.2, convergence of the results with respect to the frequency box is
  demonstrated. Section S.3 provides a comparison with CT-QMC results for
  $\lambda$. Section S.4 shows how a pseudogap develops in the present
  formalism.}\BibitemShut {Stop}%
\bibitem [{eli()}]{eliashberg}%
  \BibitemOpen
  \href@noop {} {}\bibinfo {note} {This is a first step toward the more
  involved parquet D$\Gamma$A \cite{RMPVertex}.}\BibitemShut {Stop}%
\bibitem [{Foo()}]{Footnote1}%
  \BibitemOpen
  \href@noop {} {}\bibinfo {note} {Note that a quasi two-dimensional (2D)
  system has a finite $T_{\rm c}$ if 2D layers are stacked and weakly
  interacting with each other. For a genuinely 2D system, to which the
  Mermin-Wagner theorem applies, the feedback effect of the $d$-wave pairing
  fluctuations on the self-energy needs to be taken into account, as in e.g. a
  parquet formalism \cite{Tam2013,Valli2015,Li2016}.}\BibitemShut {Stop}%
\bibitem [{\citenamefont {Sch\"afer}\ \emph {et~al.}(2017)\citenamefont
  {Sch\"afer}, \citenamefont {Katanin}, \citenamefont {Held},\ and\
  \citenamefont {Toschi}}]{Schaefer2016}%
  \BibitemOpen
  \bibfield  {author} {\bibinfo {author} {\bibfnamefont {T.}~\bibnamefont
  {Sch\"afer}}, \bibinfo {author} {\bibfnamefont {A.~A.}\ \bibnamefont
  {Katanin}}, \bibinfo {author} {\bibfnamefont {K.}~\bibnamefont {Held}}, \
  and\ \bibinfo {author} {\bibfnamefont {A.}~\bibnamefont {Toschi}},\ }\href
  {\doibase 10.1103/PhysRevLett.119.046402} {\bibfield  {journal} {\bibinfo
  {journal} {Phys. Rev. Lett.}\ }\textbf {\bibinfo {volume} {119}},\ \bibinfo
  {pages} {046402} (\bibinfo {year} {2017})}\BibitemShut {NoStop}%
\bibitem [{\citenamefont {Sakakibara}\ \emph {et~al.}(2010)\citenamefont
  {Sakakibara}, \citenamefont {Usui}, \citenamefont {Kuroki}, \citenamefont
  {Arita},\ and\ \citenamefont {Aoki}}]{sakakibara2010}%
  \BibitemOpen
  \bibfield  {author} {\bibinfo {author} {\bibfnamefont {H.}~\bibnamefont
  {Sakakibara}}, \bibinfo {author} {\bibfnamefont {H.}~\bibnamefont {Usui}},
  \bibinfo {author} {\bibfnamefont {K.}~\bibnamefont {Kuroki}}, \bibinfo
  {author} {\bibfnamefont {R.}~\bibnamefont {Arita}}, \ and\ \bibinfo {author}
  {\bibfnamefont {H.}~\bibnamefont {Aoki}},\ }\href {\doibase
  10.1103/PhysRevLett.105.057003} {\bibfield  {journal} {\bibinfo  {journal}
  {Phys. Rev. Lett.}\ }\textbf {\bibinfo {volume} {105}},\ \bibinfo {pages}
  {057003} (\bibinfo {year} {2010})}\BibitemShut {NoStop}%
\bibitem [{FLE()}]{FLEX+DMFT}%
  \BibitemOpen
  \href@noop {} {}\bibinfo {note} {Note that FLEX as supplemented by the local
  DMFT self-energy for the local part (FLEX+DMFT) also gives rather high
  $T_{\rm c}$'s \cite{Kitatani2015}.}\BibitemShut {Stop}%
\bibitem [{fre()}]{freqrange}%
  \BibitemOpen
  \href@noop {} {}\bibinfo {note} {Namely, outside the frequency box
  $[(-2n_{\rm vertex}+1)\pi/\beta \leq \nu,\nu^{\prime} \leq (2n_{\rm
  vertex}-1)\pi/\beta, (-2n_{\rm vertex}+2)\pi/\beta \leq \omega \leq (2n_{\rm
  vertex}-2)\pi/\beta ]$.}\BibitemShut {Stop}%
\bibitem [{not()}]{note-nvertex}%
  \BibitemOpen
  \href@noop {} {}\bibinfo {note} {Note that we focus on the effect of the
  vertex structure in Eq.(2), and keep Green's functions fixed here, i.e., we
  always employ, {\it unlike} in RPA, Green's functions that incorporate the
  self-energy from the full $\Gamma_{\rm m}$.}\BibitemShut {Stop}%
\bibitem [{\citenamefont {Altshuler}\ and\ \citenamefont
  {Aronov}(1985)}]{Altshuler1985}%
  \BibitemOpen
  \bibfield  {author} {\bibinfo {author} {\bibfnamefont {B.~L.}\ \bibnamefont
  {Altshuler}}\ and\ \bibinfo {author} {\bibfnamefont {A.~G.}\ \bibnamefont
  {Aronov}},\ }\href@noop {} {\emph {\bibinfo {title} {Electron-Electron
  interaction in disordered conductors}}},\ edited by\ \bibinfo {editor}
  {\bibfnamefont {A.~I.}\ \bibnamefont {Efros}}\ and\ \bibinfo {editor}
  {\bibfnamefont {M.}~\bibnamefont {Pollak}}\ (\bibinfo  {publisher} {Elsevier
  Science Publisher},\ \bibinfo {year} {1985})\BibitemShut {NoStop}%
\bibitem [{\citenamefont {Rubtsov}\ \emph {et~al.}(2012)\citenamefont
  {Rubtsov}, \citenamefont {Katsnelson},\ and\ \citenamefont
  {Lichtenstein}}]{Rubtsov12}%
  \BibitemOpen
  \bibfield  {author} {\bibinfo {author} {\bibfnamefont {A.~N.}\ \bibnamefont
  {Rubtsov}}, \bibinfo {author} {\bibfnamefont {M.~I.}\ \bibnamefont
  {Katsnelson}}, \ and\ \bibinfo {author} {\bibfnamefont {A.~I.}\ \bibnamefont
  {Lichtenstein}},\ }\href {\doibase 10.1016/j.aop.2012.01.002} {\bibfield
  {journal} {\bibinfo  {journal} {Ann. Phys.}\ }\textbf {\bibinfo {volume}
  {327}},\ \bibinfo {pages} {1320} (\bibinfo {year} {2012})}\BibitemShut
  {NoStop}%
\bibitem [{\citenamefont {Sengupta}\ and\ \citenamefont
  {Georges}(1995)}]{PhysRevB.52.10295}%
  \BibitemOpen
  \bibfield  {author} {\bibinfo {author} {\bibfnamefont {A.~M.}\ \bibnamefont
  {Sengupta}}\ and\ \bibinfo {author} {\bibfnamefont {A.}~\bibnamefont
  {Georges}},\ }\href {\doibase 10.1103/PhysRevB.52.10295} {\bibfield
  {journal} {\bibinfo  {journal} {Phys. Rev. B}\ }\textbf {\bibinfo {volume}
  {52}},\ \bibinfo {pages} {10295} (\bibinfo {year} {1995})}\BibitemShut
  {NoStop}%
\bibitem [{\citenamefont {Si}\ and\ \citenamefont {Smith}(1996)}]{Si1996}%
  \BibitemOpen
  \bibfield  {author} {\bibinfo {author} {\bibfnamefont {Q.}~\bibnamefont
  {Si}}\ and\ \bibinfo {author} {\bibfnamefont {J.~L.}\ \bibnamefont {Smith}},\
  }\href {\doibase 10.1103/PhysRevLett.77.3391} {\bibfield  {journal} {\bibinfo
   {journal} {Phys. Rev. Lett.}\ }\textbf {\bibinfo {volume} {77}},\ \bibinfo
  {pages} {3391} (\bibinfo {year} {1996})}\BibitemShut {NoStop}%
\bibitem [{\citenamefont {Smith}\ and\ \citenamefont {Si}(2000)}]{Smith00}%
  \BibitemOpen
  \bibfield  {author} {\bibinfo {author} {\bibfnamefont {J.~L.}\ \bibnamefont
  {Smith}}\ and\ \bibinfo {author} {\bibfnamefont {Q.}~\bibnamefont {Si}},\
  }\href {\doibase 10.1103/PhysRevB.61.5184} {\bibfield  {journal} {\bibinfo
  {journal} {Phys. Rev. B}\ }\textbf {\bibinfo {volume} {61}},\ \bibinfo
  {pages} {5184} (\bibinfo {year} {2000})}\BibitemShut {NoStop}%
\bibitem [{\citenamefont {Chitra}\ and\ \citenamefont
  {Kotliar}(2001)}]{Chitra01}%
  \BibitemOpen
  \bibfield  {author} {\bibinfo {author} {\bibfnamefont {R.}~\bibnamefont
  {Chitra}}\ and\ \bibinfo {author} {\bibfnamefont {G.}~\bibnamefont
  {Kotliar}},\ }\href {\doibase 10.1103/PhysRevB.63.115110} {\bibfield
  {journal} {\bibinfo  {journal} {Phys. Rev. B}\ }\textbf {\bibinfo {volume}
  {63}},\ \bibinfo {pages} {115110} (\bibinfo {year} {2001})}\BibitemShut
  {NoStop}%
\bibitem [{\citenamefont {Sun}\ and\ \citenamefont {Kotliar}(2002)}]{Sun02}%
  \BibitemOpen
  \bibfield  {author} {\bibinfo {author} {\bibfnamefont {P.}~\bibnamefont
  {Sun}}\ and\ \bibinfo {author} {\bibfnamefont {G.}~\bibnamefont {Kotliar}},\
  }\href {\doibase 10.1103/PhysRevB.66.085120} {\bibfield  {journal} {\bibinfo
  {journal} {Phys. Rev. B}\ }\textbf {\bibinfo {volume} {66}},\ \bibinfo
  {pages} {085120} (\bibinfo {year} {2002})}\BibitemShut {NoStop}%
\bibitem [{\citenamefont {Tam}\ \emph {et~al.}(2013)\citenamefont {Tam},
  \citenamefont {Fotso}, \citenamefont {Yang}, \citenamefont {Lee},
  \citenamefont {Moreno}, \citenamefont {Ramanujam},\ and\ \citenamefont
  {Jarrell}}]{Tam2013}%
  \BibitemOpen
  \bibfield  {author} {\bibinfo {author} {\bibfnamefont {K.-M.}\ \bibnamefont
  {Tam}}, \bibinfo {author} {\bibfnamefont {H.}~\bibnamefont {Fotso}}, \bibinfo
  {author} {\bibfnamefont {S.-X.}\ \bibnamefont {Yang}}, \bibinfo {author}
  {\bibfnamefont {T.-W.}\ \bibnamefont {Lee}}, \bibinfo {author} {\bibfnamefont
  {J.}~\bibnamefont {Moreno}}, \bibinfo {author} {\bibfnamefont
  {J.}~\bibnamefont {Ramanujam}}, \ and\ \bibinfo {author} {\bibfnamefont
  {M.}~\bibnamefont {Jarrell}},\ }\href {\doibase 10.1103/PhysRevE.87.013311}
  {\bibfield  {journal} {\bibinfo  {journal} {Phys. Rev. E}\ }\textbf {\bibinfo
  {volume} {87}},\ \bibinfo {pages} {013311} (\bibinfo {year}
  {2013})}\BibitemShut {NoStop}%
\bibitem [{\citenamefont {Valli}\ \emph {et~al.}(2015)\citenamefont {Valli},
  \citenamefont {Sch\"afer}, \citenamefont {Thunstr\"om}, \citenamefont
  {Rohringer}, \citenamefont {Andergassen}, \citenamefont {Sangiovanni},
  \citenamefont {Held},\ and\ \citenamefont {Toschi}}]{Valli2015}%
  \BibitemOpen
  \bibfield  {author} {\bibinfo {author} {\bibfnamefont {A.}~\bibnamefont
  {Valli}}, \bibinfo {author} {\bibfnamefont {T.}~\bibnamefont {Sch\"afer}},
  \bibinfo {author} {\bibfnamefont {P.}~\bibnamefont {Thunstr\"om}}, \bibinfo
  {author} {\bibfnamefont {G.}~\bibnamefont {Rohringer}}, \bibinfo {author}
  {\bibfnamefont {S.}~\bibnamefont {Andergassen}}, \bibinfo {author}
  {\bibfnamefont {G.}~\bibnamefont {Sangiovanni}}, \bibinfo {author}
  {\bibfnamefont {K.}~\bibnamefont {Held}}, \ and\ \bibinfo {author}
  {\bibfnamefont {A.}~\bibnamefont {Toschi}},\ }\href {\doibase
  10.1103/PhysRevB.91.115115} {\bibfield  {journal} {\bibinfo  {journal} {Phys.
  Rev. B}\ }\textbf {\bibinfo {volume} {91}},\ \bibinfo {pages} {115115}
  (\bibinfo {year} {2015})}\BibitemShut {NoStop}%
\bibitem [{\citenamefont {Li}\ \emph {et~al.}(2016)\citenamefont {Li},
  \citenamefont {Wentzell}, \citenamefont {Pudleiner}, \citenamefont
  {Thunstr\"om},\ and\ \citenamefont {Held}}]{Li2016}%
  \BibitemOpen
  \bibfield  {author} {\bibinfo {author} {\bibfnamefont {G.}~\bibnamefont
  {Li}}, \bibinfo {author} {\bibfnamefont {N.}~\bibnamefont {Wentzell}},
  \bibinfo {author} {\bibfnamefont {P.}~\bibnamefont {Pudleiner}}, \bibinfo
  {author} {\bibfnamefont {P.}~\bibnamefont {Thunstr\"om}}, \ and\ \bibinfo
  {author} {\bibfnamefont {K.}~\bibnamefont {Held}},\ }\href {\doibase
  10.1103/PhysRevB.93.165103} {\bibfield  {journal} {\bibinfo  {journal} {Phys.
  Rev. B}\ }\textbf {\bibinfo {volume} {93}},\ \bibinfo {pages} {165103}
  (\bibinfo {year} {2016})}\BibitemShut {NoStop}%
\end{thebibliography}%


\begin{thebibliography}{8}%
\makeatletter
\providecommand \@ifxundefined [1]{%
 \@ifx{#1\undefined}
}%
\providecommand \@ifnum [1]{%
 \ifnum #1\expandafter \@firstoftwo
 \else \expandafter \@secondoftwo
 \fi
}%
\providecommand \@ifx [1]{%
 \ifx #1\expandafter \@firstoftwo
 \else \expandafter \@secondoftwo
 \fi
}%
\providecommand \natexlab [1]{#1}%
\providecommand \enquote  [1]{``#1''}%
\providecommand \bibnamefont  [1]{#1}%
\providecommand \bibfnamefont [1]{#1}%
\providecommand \citenamefont [1]{#1}%
\providecommand \href@noop [0]{\@secondoftwo}%
\providecommand \href [0]{\begingroup \@sanitize@url \@href}%
\providecommand \@href[1]{\@@startlink{#1}\@@href}%
\providecommand \@@href[1]{\endgroup#1\@@endlink}%
\providecommand \@sanitize@url [0]{\catcode `\\12\catcode `\$12\catcode
  `\&12\catcode `\#12\catcode `\^12\catcode `\_12\catcode `\%12\relax}%
\providecommand \@@startlink[1]{}%
\providecommand \@@endlink[0]{}%
\providecommand \url  [0]{\begingroup\@sanitize@url \@url }%
\providecommand \@url [1]{\endgroup\@href {#1}{\urlprefix }}%
\providecommand \urlprefix  [0]{URL }%
\providecommand \Eprint [0]{\href }%
\providecommand \doibase [0]{http://dx.doi.org/}%
\providecommand \selectlanguage [0]{\@gobble}%
\providecommand \bibinfo  [0]{\@secondoftwo}%
\providecommand \bibfield  [0]{\@secondoftwo}%
\providecommand \translation [1]{[#1]}%
\providecommand \BibitemOpen [0]{}%
\providecommand \bibitemStop [0]{}%
\providecommand \bibitemNoStop [0]{.\EOS\space}%
\providecommand \EOS [0]{\spacefactor3000\relax}%
\providecommand \BibitemShut  [1]{\csname bibitem#1\endcsname}%
\let\auto@bib@innerbib\@empty
%</preamble>
\bibitem [{\citenamefont {Rohringer}\ \emph {et~al.}(2018)\citenamefont
  {Rohringer}, \citenamefont {Hafermann}, \citenamefont {Toschi}, \citenamefont
  {Katanin}, \citenamefont {Antipov}, \citenamefont {Katsnelson}, \citenamefont
  {Lichtenstein}, \citenamefont {Rubtsov},\ and\ \citenamefont
  {Held}}]{RMPvertex}%
  \BibitemOpen
  \bibfield  {author} {\bibinfo {author} {\bibfnamefont {G.}~\bibnamefont
  {Rohringer}}, \bibinfo {author} {\bibfnamefont {H.}~\bibnamefont
  {Hafermann}}, \bibinfo {author} {\bibfnamefont {A.}~\bibnamefont {Toschi}},
  \bibinfo {author} {\bibfnamefont {A.~A.}\ \bibnamefont {Katanin}}, \bibinfo
  {author} {\bibfnamefont {A.~E.}\ \bibnamefont {Antipov}}, \bibinfo {author}
  {\bibfnamefont {M.~I.}\ \bibnamefont {Katsnelson}}, \bibinfo {author}
  {\bibfnamefont {A.~I.}\ \bibnamefont {Lichtenstein}}, \bibinfo {author}
  {\bibfnamefont {A.~N.}\ \bibnamefont {Rubtsov}}, \ and\ \bibinfo {author}
  {\bibfnamefont {K.}~\bibnamefont {Held}},\ }\href {\doibase
  10.1103/RevModPhys.90.025003} {\bibfield  {journal} {\bibinfo  {journal}
  {Rev. Mod. Phys.}\ }\textbf {\bibinfo {volume} {90}},\ \bibinfo {pages}
  {025003} (\bibinfo {year} {2018})}\BibitemShut {NoStop}%
\bibitem [{\citenamefont {Rohringer}(2013)}]{Rohringer2013a}%
  \BibitemOpen
  \bibfield  {author} {\bibinfo {author} {\bibfnamefont {G.}~\bibnamefont
  {Rohringer}},\ }\emph {\bibinfo {title} {New routes towards a theoretical
  treatment of nonlocal electronic correlations}},\ \href@noop {} {Ph.D.
  thesis},\ \bibinfo  {school} {Vienna University of Technology} (\bibinfo
  {year} {2013})\BibitemShut {NoStop}%
\bibitem [{\citenamefont {Katanin}\ \emph {et~al.}(2009)\citenamefont
  {Katanin}, \citenamefont {Toschi},\ and\ \citenamefont {Held}}]{Katanin2009}%
  \BibitemOpen
  \bibfield  {author} {\bibinfo {author} {\bibfnamefont {A.~A.}\ \bibnamefont
  {Katanin}}, \bibinfo {author} {\bibfnamefont {A.}~\bibnamefont {Toschi}}, \
  and\ \bibinfo {author} {\bibfnamefont {K.}~\bibnamefont {Held}},\ }\href
  {\doibase 10.1103/PhysRevB.80.075104} {\bibfield  {journal} {\bibinfo
  {journal} {Phys. Rev. B}\ }\textbf {\bibinfo {volume} {80}},\ \bibinfo
  {pages} {075104} (\bibinfo {year} {2009})}\BibitemShut {NoStop}%
\bibitem [{\citenamefont {Werner}\ \emph {et~al.}(2006)\citenamefont {Werner},
  \citenamefont {Comanac}, \citenamefont {de' Medici}, \citenamefont {Troyer},\
  and\ \citenamefont {Millis}}]{CTHYB1}%
  \BibitemOpen
  \bibfield  {author} {\bibinfo {author} {\bibfnamefont {P.}~\bibnamefont
  {Werner}}, \bibinfo {author} {\bibfnamefont {A.}~\bibnamefont {Comanac}},
  \bibinfo {author} {\bibfnamefont {L.}~\bibnamefont {de' Medici}}, \bibinfo
  {author} {\bibfnamefont {M.}~\bibnamefont {Troyer}}, \ and\ \bibinfo {author}
  {\bibfnamefont {A.~J.}\ \bibnamefont {Millis}},\ }\href {\doibase
  10.1103/PhysRevLett.97.076405} {\bibfield  {journal} {\bibinfo  {journal}
  {Phys. Rev. Lett.}\ }\textbf {\bibinfo {volume} {97}},\ \bibinfo {pages}
  {076405} (\bibinfo {year} {2006})}\BibitemShut {NoStop}%
\bibitem [{\citenamefont {Werner}\ and\ \citenamefont {Millis}(2006)}]{CTHYB2}%
  \BibitemOpen
  \bibfield  {author} {\bibinfo {author} {\bibfnamefont {P.}~\bibnamefont
  {Werner}}\ and\ \bibinfo {author} {\bibfnamefont {A.~J.}\ \bibnamefont
  {Millis}},\ }\href {\doibase 10.1103/PhysRevB.74.155107} {\bibfield
  {journal} {\bibinfo  {journal} {Phys. Rev. B}\ }\textbf {\bibinfo {volume}
  {74}},\ \bibinfo {pages} {155107} (\bibinfo {year} {2006})}\BibitemShut
  {NoStop}%
\bibitem [{\citenamefont {Parragh}\ \emph {et~al.}(2012)\citenamefont
  {Parragh}, \citenamefont {Toschi}, \citenamefont {Held},\ and\ \citenamefont
  {Sangiovanni}}]{Parragh2012}%
  \BibitemOpen
  \bibfield  {author} {\bibinfo {author} {\bibfnamefont {N.}~\bibnamefont
  {Parragh}}, \bibinfo {author} {\bibfnamefont {A.}~\bibnamefont {Toschi}},
  \bibinfo {author} {\bibfnamefont {K.}~\bibnamefont {Held}}, \ and\ \bibinfo
  {author} {\bibfnamefont {G.}~\bibnamefont {Sangiovanni}},\ }\href {\doibase
  10.1103/PhysRevB.86.155158} {\bibfield  {journal} {\bibinfo  {journal} {Phys.
  Rev. B}\ }\textbf {\bibinfo {volume} {86}},\ \bibinfo {pages} {155158}
  (\bibinfo {year} {2012})}\BibitemShut {NoStop}%
\bibitem [{\citenamefont {Wallerberger}\ \emph {et~al.}(2019)\citenamefont
  {Wallerberger}, \citenamefont {Hausoel}, \citenamefont {Gunacker},
  \citenamefont {Kowalski}, \citenamefont {Parragh}, \citenamefont {Goth},
  \citenamefont {Held},\ and\ \citenamefont {Sangiovanni}}]{W2DYN}%
  \BibitemOpen
  \bibfield  {author} {\bibinfo {author} {\bibfnamefont {M.}~\bibnamefont
  {Wallerberger}}, \bibinfo {author} {\bibfnamefont {A.}~\bibnamefont
  {Hausoel}}, \bibinfo {author} {\bibfnamefont {P.}~\bibnamefont {Gunacker}},
  \bibinfo {author} {\bibfnamefont {A.}~\bibnamefont {Kowalski}}, \bibinfo
  {author} {\bibfnamefont {N.}~\bibnamefont {Parragh}}, \bibinfo {author}
  {\bibfnamefont {F.}~\bibnamefont {Goth}}, \bibinfo {author} {\bibfnamefont
  {K.}~\bibnamefont {Held}}, \ and\ \bibinfo {author} {\bibfnamefont
  {G.}~\bibnamefont {Sangiovanni}},\ }\href@noop {} {\bibfield  {journal}
  {\bibinfo  {journal} {Computer Physics Communications}\ }\textbf {\bibinfo
  {volume} {235}},\ \bibinfo {pages} {388 } (\bibinfo {year}
  {2019})}\BibitemShut {NoStop}%
\bibitem [{\citenamefont {Sch\"afer}\ \emph {et~al.}(2017)\citenamefont
  {Sch\"afer}, \citenamefont {Katanin}, \citenamefont {Held},\ and\
  \citenamefont {Toschi}}]{Schaefer2016}%
  \BibitemOpen
  \bibfield  {author} {\bibinfo {author} {\bibfnamefont {T.}~\bibnamefont
  {Sch\"afer}}, \bibinfo {author} {\bibfnamefont {A.~A.}\ \bibnamefont
  {Katanin}}, \bibinfo {author} {\bibfnamefont {K.}~\bibnamefont {Held}}, \
  and\ \bibinfo {author} {\bibfnamefont {A.}~\bibnamefont {Toschi}},\ }\href
  {\doibase 10.1103/PhysRevLett.119.046402} {\bibfield  {journal} {\bibinfo
  {journal} {Phys. Rev. Lett.}\ }\textbf {\bibinfo {volume} {119}},\ \bibinfo
  {pages} {046402} (\bibinfo {year} {2017})}\BibitemShut {NoStop}%
\end{thebibliography}%

\newcommand{\PR}[3]{Phys. Rev. \textbf{#1},#2 (#3)}
\newcommand{\PRL}[3]{Phys. Rev. Lett. \textbf{#1},#2 (#3)}
\newcommand{\PRA}[3]{Phys. Rev. A \textbf{#1}, #2 (#3)}
\newcommand{\PRB}[3]{Phys. Rev. B \textbf{#1}, #2 (#3)}
\newcommand{\JPSJ}[3]{J. Phys. Soc. Jpn. \textbf{#1}, #2 (#3)}
\newcommand{\arxiv}[1]{arXiv:#1}
\newcommand{\RMP}[3]{Rev. Mod. Phys. \textbf{#1}, #2 (#3)}

\end{document}

\title{Supplemental materials to \\
``Why the critical temperature of high-$T_{\rm c}$ cuprate superconductors is so low: \\
The importance of the dynamical vertex structure"}

\author{Motoharu Kitatani}
\affiliation{Institute of Solid State Physics, Vienna University of Technology, A-1040 Vienna, Austria}
\affiliation{Department of Physics, University of Tokyo, Hongo, Tokyo 113-0033, Japan}

\author{Thomas Sch\"{a}fer}
\affiliation{Institute of Solid State Physics, Vienna University of Technology, A-1040 Vienna, Austria}
\affiliation{Coll\`{e}ge de France, 11 place Marcelin Berthelot, 75005 Paris, France}
\affiliation{Centre de Physique Th\'{e}orique, \'{E}cole Polytechnique, CNRS, route de Saclay, 91128 Palaiseau, France}

\author{Hideo Aoki}
\affiliation{Department of Physics, University of Tokyo, Hongo, Tokyo 113-0033, Japan}
\affiliation{Electronics and Photonics Research Institute,
Advanced Industrial Science and Technology (AIST), Tsukuba, Ibaraki 305-8568, Japan}

\author{Karsten Held}
\affiliation{Institute of Solid State Physics, Vienna University of Technology, A-1040 Vienna, Austria}

\begin{abstract}
\end{abstract}

\maketitle

\section{Calculation of the pairing vertex $\Gamma{\rm pp}$}
In this section, we explain how to calculate the pairing interaction $\Gamma_{\rm pp}$
within the ladder D$\Gamma$A formalism. 
We can fix one bosonic frequency ($\omega$) to treat a two-particle quantity as a matrix that 
depends on two fermionic frequencies ($\nu,\nu^{\prime}$).
The nonlocal two-particle vertex 
$F_{\rm r}$ is then calculated in the ladder expansion through the matrix equation
based on local irreducible matrix $\Gamma_{\rm r}$,
\begin{equation}
	F_{\rm r}= \Gamma_{\rm r} - \Gamma_{\rm r} \chi_0 F_{\rm r},
\end{equation}
with ${\rm r}={\rm d,m}$ 
denoting the density (d) or magnetic (m) channel, 
and $\chi_0$ is the bare bubble susceptibility. 
The density and magnetic vertex $\Gamma_{\rm d/m}$ are obtained by 
$\Gamma_{\rm d/m}=\Gamma_{\uparrow \uparrow} \pm \Gamma_{\uparrow \downarrow}$,
whose particle-hole convention is displayed in Fig.~S.1(a).
Let us express $\Gamma_{\rm r}$ as $\Gamma_{\rm r} = {\cal U}_{\rm r} + \delta \Gamma_{\rm r}$ as in the main text,
where ${\cal U}_{\rm r}$ is the matrix with all the elements being $U_{\rm d}=U$ and $U_{\rm m}=-U$ 
for ${\rm r} = {\rm d}$ and {\rm m}, respectively \cite{RMPvertex}.
Then the Bethe Salpeter equation reads
\begin{align}
F_{\rm r} =& ( {\cal U}_{\rm r} + \delta \Gamma_{\rm r}) \sum_{n=0}^{\infty} [-\chi_0 ( {\cal U}_{\rm r} +\delta \Gamma_{\rm r})]^n \nonumber \\
	=& ({\cal U}_{\rm r} + \delta \Gamma_{\rm r}) \sum_{n=0}^{\infty} (-\chi_0 \delta \Gamma_{\rm r})^n %\nonumber \\
	+ ( {\cal U}_{\rm r} + \delta \Gamma_{\rm r}) \left[ \sum_{n=0}^{\infty} (-\chi_0 \delta \Gamma_{\rm r})^n \right]
		(-\chi_0 {\cal U}_{\rm r} ) \left[ \sum_{n=0}^{\infty} (-\chi_0 \delta \Gamma_{\rm r})^n \right] \nonumber \\
	&+ ( {\cal U}_{\rm r} + \delta \Gamma_{\rm r}) \left[ \sum_{n=0}^{\infty} (-\chi_0 \delta \Gamma_{\rm r})^n \right]
		(-\chi_0 {\cal U}_{\rm r}) \left[ \sum_{n=0}^{\infty} (-\chi_0 \delta \Gamma_{\rm r})^n \right]
		(-\chi_0 {\cal U}_{\rm r}) \left[ \sum_{n=0}^{\infty} (-\chi_0 \delta \Gamma_{\rm r})^n \right] + \cdots.
\end{align}

If we collect the terms summed over $\nu,\nu^{\prime}$ defined as
\begin{equation}
	{\bar{\chi}}_{\rm r} \equiv \sum_{\nu,\nu^{\prime}} \left[ \sum_{n=0}^{\infty} (-\chi_0 \delta \Gamma_{\rm r})^n \right] \chi_0,
\label{eq:chi-bar}
\end{equation}
which is feasible due to the simple structure of ${\cal U}_{\rm r}$,
$F_{\rm r}$ becomes
\begin{align}
F_{\rm r} &= {\cal U}_{\rm r}(1-{\bar{\chi}}_{\rm r}U_{\rm r}+
      {\bar{\chi}}_{\rm r}U_{\rm r}{\bar{\chi}}_{\rm r}U_{\rm r}+ \cdots) 
      \left[\sum_{n=0}^{\infty} (-\chi_0 \delta \Gamma_{\rm r})^n \right] \nonumber \\
	+& \delta \Gamma_{\rm r} \left[ \sum_{n=0}^{\infty}(-\chi_0 \delta \Gamma_{\rm r})^n \right]
	 (-\chi_0) {\cal U}_{\rm r}(1-{\bar{\chi}}_{\rm r}U_{\rm r}+{\bar{\chi}}_{\rm r}U_{\rm r}{\bar{\chi}}_{\rm r}U_{\rm r}+ 
	  \cdots) \left[\sum_{n=0}^{\infty} (-\chi_0 \delta \Gamma_{\rm r})^n \right]
	+ \delta \Gamma_{\rm r} \left[ \sum_{n=0}^{\infty}(-\chi_0 \delta \Gamma_{\rm r})^n \right] \nonumber \\
	&= \left[ \sum_{n=0}^{\infty}(-\delta \Gamma_{\rm r} \chi_0)^n \right] 
	  {\cal U}_{\rm r} \left( 1 - \frac{U_{\rm r} {\bar{\chi}}_{\rm r}}{1+U_{\rm r}{\bar{\chi}}_{\rm r}} \right)
	   \left[ \sum_{n=0}^{\infty}(-\chi_0 \delta \Gamma_{\rm r})^n \right] %\nonumber \\
	   + \delta \Gamma_{\rm r} \left[ \sum_{n=0}^{\infty}(-\chi_0 \delta \Gamma_{\rm r})^n \right] \nonumber \\
	&= \left[ \sum_{n=0}^{\infty}(-\delta \Gamma_{\rm r} \chi_0)^n \right] 
	  {\cal U}_{\rm r} \left( 1 - U_{\rm r} \chi_{\rm r} \right) 
	   \left[ \sum_{n=0}^{\infty}(-\chi_0 \delta \Gamma_{\rm r})^n \right]
	   + \delta \Gamma_{\rm r} \left[ \sum_{n=0}^{\infty}(-\chi_0 \delta \Gamma_{\rm r})^n \right], %\nonumber \\
\end{align}
where $\chi_{\rm r}= {\bar{\chi}}_{\rm r}/(1+U_{\rm r}{\bar{\chi}}_{\rm r})$ is the physical susceptibility, 
and we have used that $1+ \delta \Gamma_{\rm r} \sum_{n=0}^{\infty}(-\chi_0 \delta \Gamma_{\rm r})^n (-\chi_0) = \sum_{n=0}^{\infty}(-\delta \Gamma_{\rm r}\chi_0)^n$.
The above equation is a generalization of the discussion in the main text; it explicitly shows that
$F_{\rm r}$ reduces to RPA in the limit $\delta \Gamma_{\rm r} \rightarrow 0$. 
If we introduce a finite $\delta \Gamma_{\rm r}$, the physical susceptibility is modified through Eq.~({\ref{eq:chi-bar}}).
We can see that the vertex correction $\delta \Gamma$ affects not only the physical susceptibility (that depends only on $\omega$)
but also the vertex structure itself (which depends on $\nu,\nu^{\prime},\omega$).
While we now employ $\delta \Gamma_{\rm r}$ following the main text, 
we can alternatively express $F_{\rm r}$ without infinite summation as 
\begin{equation}
F_{\rm r}^{\nu \nu^{\prime}}=(\chi_0^{\nu})^{-1}
\left[ \delta_{\nu,\nu^{\prime}}-\chi_{\rm r}^{*,\nu\nu^{\prime}}(\chi_0^{\nu^{\prime}})^{-1} \right]
+ U_{\rm r} (1-U_{\rm r} \chi_{\rm r}) \gamma_{\rm r}^{\nu} \gamma_{\rm r}^{\nu^{\prime}},
\end{equation}
where $(\chi^{*})^{-1} \equiv \chi_0^{-1} + \delta \Gamma_{\rm r}$ and 
$\gamma_{\rm r}^{\nu} \equiv (\chi_0^{\nu})^{-1} \sum_{\nu^{\prime}} \chi_{\rm r}^{*,\nu\nu^{\prime}}$. 
This is the same expression as previously employed in ladder D$\Gamma$A (for details see \cite{Rohringer2013a}), 
and we have used this formulation in the actual calculations.
As in the self-energy calculation, we consider a Moriyaesque $\lambda$-correction for $F$ by replacing the physical $\chi_{\rm r}$
as \cite{Rohringer2013a,RMPvertex,Katanin2009}
\begin{equation}
\chi_{\rm r}^{\lambda} \equiv (\chi_{\rm r}^{-1} + \lambda_{\rm r})^{-1}.
\end{equation}

In the ladder D$\Gamma$A, 
the two-particle vertex $F_{\uparrow \downarrow}$ is related to 
the ladder-expanded vertex $F_{\rm r}$ through
\begin{align}
F_{{\rm ladder},\uparrow \downarrow} (k,k^{\prime},q) =& 
\frac{1}{2}[F_{{\rm d},{\bm{q}}}(\nu,\nu^{\prime},\omega) - F_{{\rm m},{\bm{q}}}(\nu,\nu^{\prime},\omega)] \notag\\
&- F_{{\rm m},{\bm{k^{\prime}-k}}}(\nu,\nu+\omega,\nu^{\prime}-\nu) 
-F_{{\rm loc},\uparrow \downarrow}(\nu,\nu^{\prime},\omega).
\end{align}
Here, both the particle-hole and the transversal particle-hole channels are considered with the local double-counting terms subtracted \cite{Rohringer2013a,RMPvertex}, 
%- \frac{1}{2}[ F_{{\rm loc},d}(\omega_n,\omega_{n^{\prime}},\omega_m) 
%-F_{{\rm loc},m}(\omega_n,\omega_{n^{\prime}},\omega_m)],
and the pairing vertex is calculated as 
\begin{equation}
\Gamma_{\rm pp}(k,k^{\prime},q=0) = F_{\rm ladder}(k^{\prime},-k,k-k^{\prime}) - \Phi_{\rm loc,pp}(k,k^{\prime},q=0) 
\end{equation}
for which the particle-particle convention as displayed in Fig.~S.1(b) is employed.
For singlet pairing, for which $\Delta(k) =\Delta(-k)$, we can use $\Gamma^{\rm s}_{\rm pp}$ defined as 
\begin{align}
\Gamma_{{\rm pp}}^{\rm s}(k,k^{\prime},q=0) 
\equiv \Gamma_{{\rm pp},\bm{Q}=\bm{k}-\bm{k}^{\prime}}^{\rm s}(\nu,\nu^{\prime})
=&
\frac{1}{2} F_{{\rm d},{\bm{k-k^{\prime}}}}(\nu^{\prime},-\nu,(\nu-\nu^{\prime}))
-\frac{3}{2} F_{{\rm m},{\bm{k-k^{\prime}}}}(\nu^{\prime},-\nu,(\nu-\nu^{\prime})) \notag\\
&-F_{{\rm loc},\uparrow \downarrow}(\nu^{\prime},-\nu,(\nu-\nu^{\prime}))
-\Phi_{{\rm loc},{\rm pp}} (\nu,\nu^{\prime},\omega=0),
\end{align}
which is the same as $\Gamma_{\rm pp}$ for singlet eigenvectors 
if we take the symmetric component under the sign change of $k$ ($k \leftrightarrow -k$).
For triplet pairing, for which $\Delta(k) = -\Delta(-k)$, we can similarly 
use $\Gamma^{\rm t}_{\rm pp}$ defined as
\begin{align}
\Gamma_{\rm pp}^{\rm t}(k,k^{\prime},q=0) 
\equiv \Gamma_{{\rm pp},\bm{Q}=\bm{k}-\bm{k}^{\prime}}^{\rm t}(\nu,\nu^{\prime})
=&
\frac{1}{2} F_{{\rm d},{\bm{k-k^{\prime}}}}(\nu^{\prime},-\nu,(\nu-\nu^{\prime}))
+\frac{1}{2} F_{{\rm m},{\bm{k-k^{\prime}}}}(\nu^{\prime},-\nu,(\nu-\nu^{\prime})) \notag\\
&-F_{{\rm loc},\uparrow \downarrow}(\nu^{\prime},-\nu,(\nu-\nu^{\prime}))
-\Phi_{{\rm loc},{\rm pp}}(\nu,\nu^{\prime},\omega=0).
\end{align}

\begin{figure}[t]
\begin{centering}
\includegraphics[width=0.75\columnwidth]{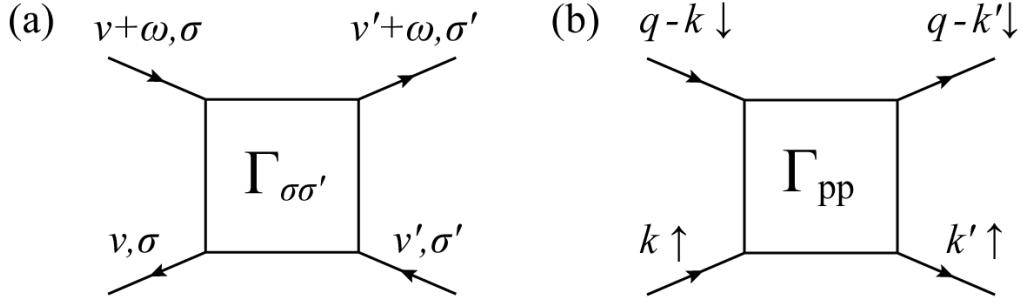}
\par\end{centering}
\caption{
(a) Local particle-hole irreducible vertex $\Gamma_{\sigma \sigma^{\prime}}(\nu,\nu^{\prime},\omega)$ and
(b) particle-particle irreducible vertex $\Gamma_{\rm pp}(k,k^{\prime},q)$.
}
\label{fig:diags}
\end{figure}

\section{Convergence against frequency range of the vertex}
Due to the heavy computational cost, we can only consider a limited range 
of frequency ($n_{\rm core} \approx 160$ positive frequencies) 
for the vertex, while superconductivity occurs at quite low temperatures 
which corresponds to a fine frequency grid. 
Therefore, it is a challenge to obtain well-converged results against the frequency range 
of the local two-particle quantities in DMFT. Here, instead of simply solving the Bethe-Salpeter equation
for such a limited frequency range, 
we consider a wide frequency range ($n_{\rm outer}$ points on the positive side). 
For the outer frequencies, we use the bare $ {U}_{\rm r}=\pm U$ instead of $\Gamma_{\rm ph}$
(which is restricted to $n_{\rm core}$ positive frequencies). 
This supplementation of $\Gamma_{\rm ph}$ by  $ {U}_{\rm r}$ is done twice: 
once for calculating $\Gamma_{\rm ph}$ from the impurity susceptibility and
again when solving the Bethe-Salpeter equation.
Figure S.2 shows the dependence of the result on the inner range $n_{\rm core}$ (the range for the two-particle data in DMFT) 
and on the outer range $n_{\rm outer}$ (the range for the bare $U$ contribution).
We can see that the $n_{\rm core}$ dependence quickly converges if we take into 
account a wide range of the bare $U$ contribution. 
While convergence with respect to $n_{\rm outer}$ is not fast, we can treat a large frequency range for the bare $U$ contribution 
without much computational costs 
(note the computational bottleneck is the exact diagonalization (ED) calculation of the local vertex). 
In the calculation shown in Fig.~1 in the main text, we take $n_{\rm outer}$=1024 and up to $n_{\rm core}=100$ for the lowest temperature ($T=0.01t$), and 
for Figs.~2 and 3 we take $n_{\rm outer}=1024$ and $n_{\rm core}=120$.

\begin{figure}[t]
\begin{centering}
\includegraphics[width=1\columnwidth]{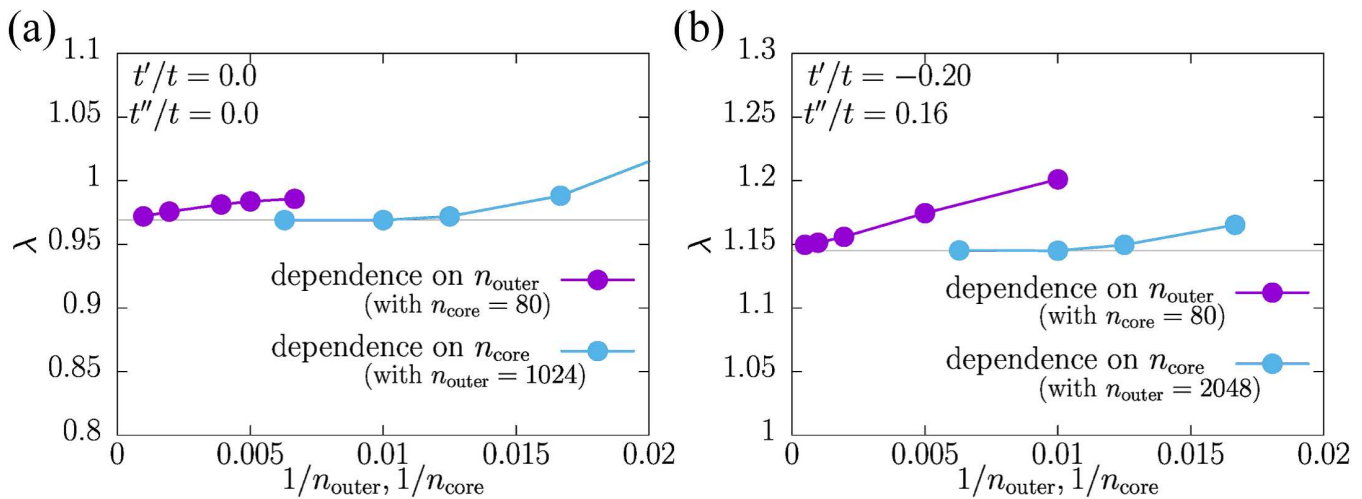}
\par\end{centering}
\caption{%(Color online)
Dependence of the $d$-wave superconductivity eigenvalue on $1/n_{\rm core}$ 
for a fixed $n_{\rm outer}$, or 
$1/n_{\rm outer}$ for a fixed $n_{\rm core}$ for 
(a) $n=0.825,t^{\prime}=t^{\prime \prime}=0$, and (b) $n=0.90, t^{\prime}/t=-0.20,t^{\prime \prime}/t=0.16$ with $U/t=6.0,T=0.01t$. 
These parameters correspond approximately to the optimal doping region in Fig.~1(a),(b) in the main text.  
Horizontal lines are guides to the eye for the (converged) $\lambda$ value at $n_{\rm core}= 159$ for $n_{\rm outer}=1024$ (left) and $n_{\rm outer}=2048$ (right).}
\label{fig:diags}
\end{figure}

\section{Comparison with CT-QMC results}
Here, we show the comparison with the eigenvalues $\lambda$ that are obtained with 
continuous-time quantum Monte Carlo (CT-QMC) \cite{CTHYB1,CTHYB2,Parragh2012,W2DYN} as an impurity solver.
The result in Fig.S.3 demonstrates that the ED discretization error hardly affects the $\lambda$ values presented in the present paper.
Note that the error which comes from using different $(n_{\rm core}, n_{\rm outer})$ 
for ED and CT-QMC is smaller than the size of symbols. We have also checked the calculated $\Sigma(k)$ 
agrees with CT-QMC result.

\begin{figure}[h]
\begin{centering}
\includegraphics[width=0.4\columnwidth]{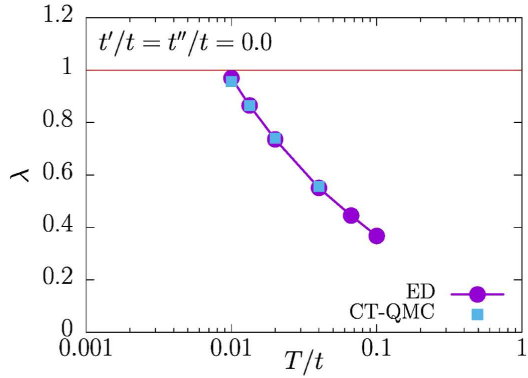}
\par\end{centering}
\caption{
Comparison of results for the $d$-wave superconductivity eigenvalue $\lambda$ between two impurity solvers: 
CT-QMC (with $n_{\rm outer}=400$ and $n_{\rm core}=50$ for $T \leq 0.013t$, and $n_{\rm core}=80$ for $T=0.010t$) and ED 
(with $n_{\rm outer}=1024$ and $n_{\rm core}=50$ for $T \leq 0.02t$, $n_{\rm core}=80$ for $T=0.013t$, and $n_{\rm core}=100$ for $T=0.010t$). 
Here we take $U/t=6.0, t^{\prime}=t^{\prime \prime}=0$, and $n=0.825$
which correspond approximately to the optimal doping region in Fig.~1(a) in the main text.}
\label{fig:diags}
\end{figure}

\begin{figure}
\begin{centering}
\includegraphics[width=0.8\columnwidth]{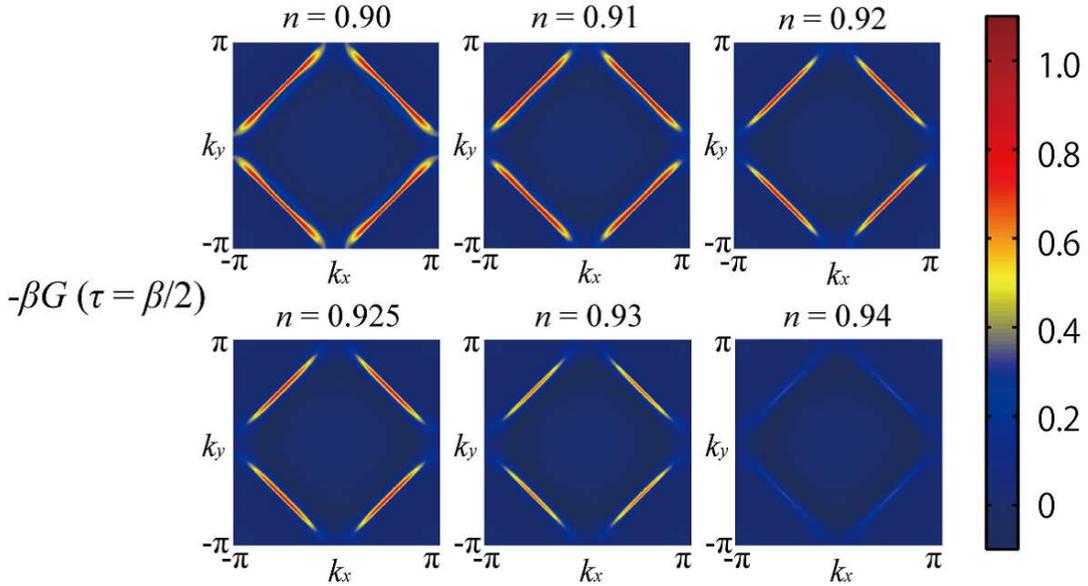}
\par\end{centering}
\caption{
Momentum dependence of $G({\bm k},\tau=\beta/2)$ for 
the fillings indicated %$n=0.90, 0.91, 0.92, 0.925, 0.93, 0.94$ 
at $U=6t, t'=t''=0, T/t=0.02.$
}
\label{fig:G-beta-half}
\end{figure}

\section{Momentum dependence of the spectral weight}
In this section, we discuss spectra in the D$\Gamma$A calculations in more detail, to specifically look at how a pseudogap develops. 
First in Fig.~S.4, instead of showing the spectral weight obtained with analytical continuation, we present Green's function $G$ at imaginary time $\tau=\beta/2$
\begin{equation}
-{\beta} G({\bm k},\tau=\beta/2) = \pi \int A({\bm k},\omega) \left[ \frac{\beta}{2 \pi} \frac{1}{{\rm cosh}(\frac{\beta \omega}{2})} \right] {\rm d}\omega,
\end{equation}
which reflects the spectral weight $A({\bm k},\omega=0)$ with a finite temperature blurring and has the advantage that no analytical continuation is needed.
We can see, in Fig.~S.4 for various   fillings, a strong reduction of the spectral weight around the 
anti-nodal directions along ${\bm k} = (\pm\pi,0), (0,\pm\pi)$. 
Accordingly, the Fermi surface shrinks into Fermi arcs as we approach 
the half filling, as observed in experiments for cuprates.  
In Fig.S.5 we note that in this region, $|G({\bm k},\omega_n=\pi/\beta)|$ has quite a different structure from the spectral weight Eq.(S.10),
while ${\rm Im}G({\bm k},\omega_n=\pi/\beta)$ is similar to $\beta G({\bm k},\tau=\beta/2)$ 
but with larger temperature blurring.

Let us next show in Fig.~S.6 the analytically continued density of states $\rho(\omega) = \sum_{\bm k} A({\bm k},\omega)$
as obtained from a  Pad\'{e} fit, at a higher temperature ($T=0.1t$) where the Pad\'{e} fit works more reliably.
Here, we can see a salient pseudogap structure toward  half-filling ($n \ge 0.90$) as well as the lower and upper Hubbard band structure.

These demonstrate that the D$\Gamma$A approach describes a pseudogap structure 
along with Hubbard bands for the parameter region studied here. 
The method also describes that antiferromagnetic correlations (and ordering) fade away when doping puts the system 
away from half filling, see Fig.~2~(d) in the main text and Ref.~\cite{Schaefer2016}.

\begin{figure}
\begin{centering}
\includegraphics[width=0.8\columnwidth]{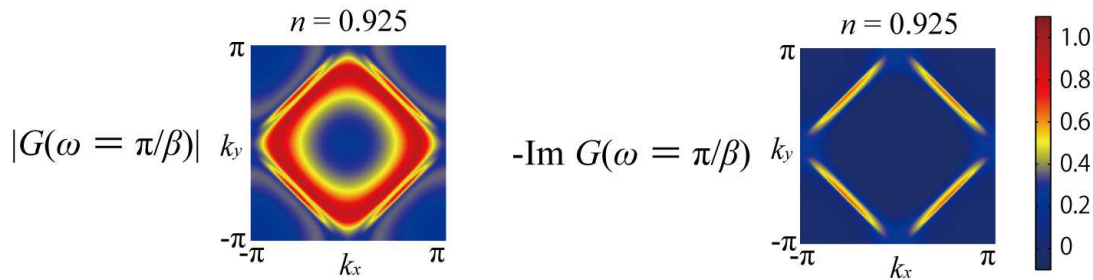}
\par\end{centering}
\caption{Green's function 
$|G({\bm k},\omega_n=\pi/\beta)|$ and $-{\rm Im}G({\bm k},\omega_n=\pi/\beta)$ 
plotted against momentum for 
a filling $n=0.925$. Other parameters and color codes are the same as in  Fig.~S.4.  Note a difference in the color code from Fig.~2 (c) in the main text.
}
\label{fig:ABSG}
\end{figure}

\begin{figure}
\begin{centering}
\includegraphics[width=1.0\columnwidth]{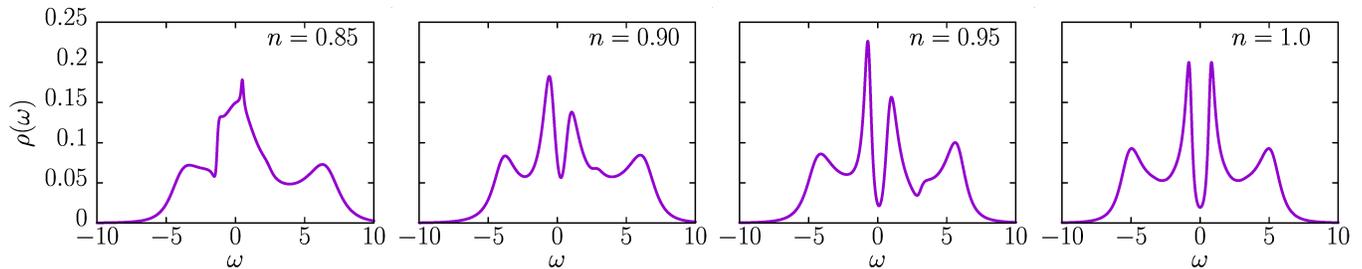}
\par\end{centering}
\caption{Densities of states for the fillings indicated 
at $U=6t, t'=t''=0, T/t=0.1.$}
\end{figure}

\newpage
\bibliography{main,addrefs}

\newcommand{\PR}[3]{Phys. Rev. \textbf{#1},#2 (#3)}
\newcommand{\PRL}[3]{Phys. Rev. Lett. \textbf{#1},#2 (#3)}
\newcommand{\PRA}[3]{Phys. Rev. A \textbf{#1}, #2 (#3)}
\newcommand{\PRB}[3]{Phys. Rev. B \textbf{#1}, #2 (#3)}
\newcommand{\JPSJ}[3]{J. Phys. Soc. Jpn. \textbf{#1}, #2 (#3)}
\newcommand{\arxiv}[1]{arXiv:#1}
\newcommand{\RMP}[3]{Rev. Mod. Phys. \textbf{#1}, #2 (#3)}